\documentclass[11pt]{article}

\usepackage{type1cm}        
\usepackage{makeidx}         
\usepackage{graphicx}        
\usepackage{multicol}        
\usepackage[bottom]{footmisc}

\usepackage{newtxtext}       %
\usepackage[varvw]{newtxmath}       

\usepackage{fullpage}

\usepackage{cite}
\usepackage{caption}
\usepackage{subcaption}
\usepackage{titlesec}
\usepackage{amsmath}
\usepackage{tabularx}
\usepackage{bm}
\usepackage{color}
\usepackage{url}
\usepackage{bigfoot}
\usepackage{multirow}
\usepackage{setspace}

\usepackage{placeins}
\usepackage{inputenc}
\usepackage{ulem}
\usepackage{upgreek}

\def\##1{{\bf #1}}
\def\=#1{\underline{\underline #1}}
\def\^#1{{\rR}eve{#1}}
\def\`#1{{#1^\prime}}
\def\:#1{#1^{\prime\prime}}

\def\.{\mbox{ \tiny{$^\bullet$} }}

\def\ux{\hat{\#x}}
\def\uy{\hat{\#y}}
\def\uz{\hat{\#z}}

\def\muo{\mu_{\scriptscriptstyle 0}}
\def\epso{\eps_{\scriptscriptstyle 0}}

\def\lambdao{\lambda_{\scriptscriptstyle 0}}
\def\ko{k_{\scriptscriptstyle 0}}

\def\eps{\varepsilon}
\def\epsa{\eps_{\rm a}}
\def\epsb{\eps_{\rm b}}
\def\epsc{\eps_{\rm c}}

\def\le{\left(}
\def\ri{\right)}
\def\les{\left[}
\def\ris{\right]}
\def\lec{\left\{}
\def\ric{\right\}}

\def\deg{^\circ}

\def\aL{a_{\rm L}}
\def\aR{a_{\rm R}}
\def\rL{r_{\rm L}}
\def\rR{r_{\rm R}}
\def\tL{t_{\rm L}}
\def\tR{t_{\rm R}}

 \def\TL{T_{\rm L}}
\def\TR{T_{\rm R}}
 \def\RL{R_{\rm L}}
\def\RR{R_{\rm R}}

\def\as{a_{\rm s}}
\def\ap{a_{\rm p}}
\def\rs{r_{\rm s}}
\def\rp{r_{\rm p}}
\def\ts{t_{\rm s}}
\def\tp{t_{\rm p}}

\def\Rs{R_{\rm s}}
\def\Rp{R_{\rm p}}
\def\Ts{T_{\rm s}}
\def\Tp{T_{\rm p}}

\def\thetainc{\theta_{\rm inc}}

\def\sinczero{s_0^{\rm inc}}
\def\sincone{s_1^{\rm inc}}
\def\sinctwo{s_2^{\rm inc}}
\def\sincthree{s_3^{\rm inc}}

\def\alphainc{\alpha^{\rm inc}}
\def\betainc{\beta^{\rm inc}}

\def\bphit{\les{\underline \phi}\ris}
\def\bphitinc{\les{\underline \phi}^{\rm inc}\ris}

\def\srefzero{s_0^{\rm ref}}
\def\srefone{s_1^{\rm ref}}
\def\sreftwo{s_2^{\rm ref}}
\def\srefthree{s_3^{\rm re}}

\def\alpharef{\alpha^{\rm ref}}
\def\betaref{\beta^{\rm ref}}

\def\bphitref{\les{\underline \phi}^{\rm ref}\ris}

\def\alphatra{\alpha^{\rm tr}}
\def\betatra{\beta^{\rm tr}}

\def\bphittra{\les{\underline \phi}^{\rm tr}\ris}

\def\bPhi{\Phi}

\def\bPhiref{\Phi^{\rm ref}}

\def\bPhirefL{ \Phi_{\rm L}^{\rm ref}}
\def\bPhirefR{\Phi_{\rm R}^{\rm ref}}

\def\bPhitra{\Phi^{\rm tr}}

\def\bPhitraL{ \Phi_{\rm L}^{\rm tr}}
\def\bPhitraR{\Phi_{\rm R}^{\rm tr}}

\def\sp{\#s}
\def\pinc{\#p_+}
\def\pref{\#p_-}

\def\Ei{\#E^{\rm inc}(\#r)}

\def\Er{\#E^{\rm ref}(\#r)}

\def\Et{\#E^{\rm tr}(\#r)}

\def\Nlyr{\tilde{N}}

\begin{document}

\begin{center} 

\textbf{\Large Geometric Phase and Nanoscale Architected Morphology of Reusch Piles}\\

\textit{Akhlesh Lakhtakia}\\

Department of Engineering
Science and Mechanics, The Pennsylvania State University, \\ University Park, PA, USA \\
and\\
School of Mathematics, University of Edinburgh, Edinburgh EH9 3FD, UK\\

\abstract{The geometric phase has acquired interest for optical devices such as achromatic phase shifters, spatial light modulators, frequency shifters, and planar lenses for wavefront engineering.  Numerical work with  Reusch piles with a large number of layers per period
suggests that the geometric phase of the reflected/transmitted plane wave may be sensitive to nanoscale
morphological differences, and the geometric-phase spectrum could contain signatures of  morphological details that  depend
on the fabrication technique. }
\vspace{5mm}


 \end{center}

\section{Introduction}\label{sec:intro} 

\subsection{Poincar\'e spinor}
\begin{figure}[ht]
	\begin{center}
			\includegraphics[width=0.2\linewidth]{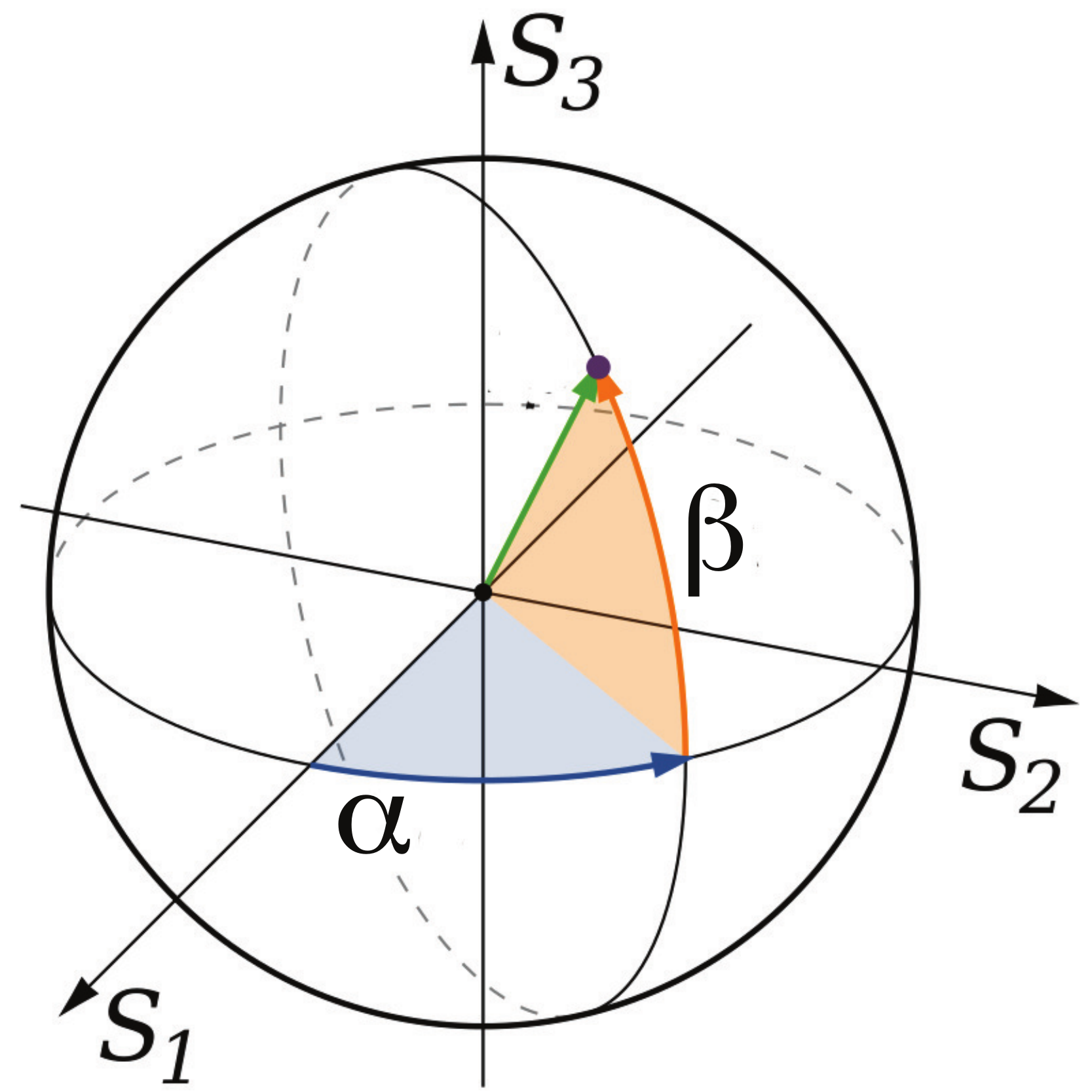}
	\end{center}
	\caption{Longitude $\alpha\in[0,2\pi)$ and latitude
	$\beta\in[-\pi/2,\pi/2]$ that identify a plane wave on the Poincar\'e sphere.}
	\label{Poincare-sphere}
\end{figure}

Any uniform plane wave propagating in free space can be represented as a point on the surface of the Poincar\'e sphere
$s_1^2+s_2^2+s_3^2=s_0^2$, where $s_0$, $s_{1}$, $s_2$, and $s_3$ are the four Stokes parameters. The plane wave's location
is identified by the longitude $\alpha\in[0,2\pi)$ and the latitude $\beta\in[-\pi/2,\pi/2]$ defined through the relations
\begin{equation}
\label{def-alphabeta}
\left.\begin{array}{l}
s_1=s_0\, \cos\beta\, \cos\alpha
\\[5pt]
s_2=s_0\, \cos\beta\, \sin\alpha
\\[5pt]
s_3=s_0\, \sin\beta 
\end{array}
\right\}\,,
\end{equation}
as shown in Fig.~\ref{Poincare-sphere}. These two angles appear in the  Poincar\'e spinor
\begin{equation}
\label{def-PS}
\bphit=\les
\begin{array}{c}
\cos\left(\frac{\pi}{4}-\frac{\beta}{2}\right)
\\[5pt]
\sin\left(\frac{\pi}{4}-\frac{\beta}{2}\right)\exp\left(i\alpha\right)
\end{array}
\ris\,,
\end{equation}
where $i=\sqrt{-1}$.

 \subsection{Geometric phase}
 
With respect to a plane wave labeled ``1",
the geometric phase of a plane wave labeled ``2"  is defined
as the angle
\begin{equation}
\Phi_{21}={\rm Arg}\lec {\les{\underline \phi}_1\ris}^\dag\cdot{\les{\underline \phi}_2\ris}\ric\,,
\end{equation}
where $^\dag$ denotes the conjugate transpose. The geometric phase is a measure of dissimilarity of two plane waves of the same frequency, even when propagating in the same direction, as was pointed out by Pancharatnam in 1956 \cite{Pancha}. The profound role of geometric phase in classical as well as quantum physics was recognized within the next three decades  \cite{SWbook} and it continues to both fascinate researchers \cite{CLBNGK,Citro} and find applications \cite{Brasselet2017,Kobashi2016,metasurface,Jisha}.

Optical experiments on plane-wave transmission through a columnar thin film \cite{HWbook}
and a chiral sculptured thin film \cite{STFbook} have revealed that the geometric phase of the transmitted plane wave depends on the morphology of the thin film \cite{Das}. Thin-film morphology can be architected using a variety of fabrication techniques including
physical vapor deposition \cite{STFbook,Martin}, holographic lithography \cite{Juodkazis},
 direct laser writing \cite{Thiel}, electron tomography \cite{KHMP}, hydrothermal method \cite{Ding},
 and self assembly \cite{Kelly,Yeom}. Each technique may produce thin films with similar plane-wave reflectance and transmittance characteristics, but the geometric phase of the reflected and/or transmitted
 plane wave could contain a  signature of the fabrication technique.
 
 \subsection{Reusch piles}\label{sec:RP}
 An experimental test of the foregoing proposition requires resources presently unavailable to me. So, I chose to establish the rudiments of this proposition theoretically. Chiral sculptured thin films may be considered to be \textit{finely chiral} Reusch piles \cite{ambi}. Conceived in 1869 \cite{Reusch}, a Reusch pile
 is a periodic multilayer comprising layers of the same
homogeneous, uniaxial dielectric material,  such that the optic axis in each layer
is rotated about the thickness direction (designated to be parallel to the $z$ axis
of a Cartesian coordinate system) with respect to the optic axis in the
previous layer by a fixed angle $\Delta\xi$. A delightful artistic interpretation
of Reusch piles is available from the sculpture \textit{Seque} created by Anne
Huibregtse \cite{Zaidi}.

A Reusch pile
preferentially reflects light of one circular-polarization state in multiple spectral
regimes \cite{Joly2},  preserving the
circular-polarization state of the incident plane wave \cite{Joly4}.
These spectral regimes, called \textit{circular Bragg regimes} \cite{FLaop}, depend on both the number of layers
in one period of the Reusch pile and the dependencies of the constitutive parameters
on the frequency \cite{ambi}.

The following classification of Reusch piles was devised in 2004 \cite{ambi}: Suppose that the period $P$
of a Reusch pile is made up of $2q$, $q\geq2$, identical layers stacked along the
$z$ axis. Then the angular offset of the optic axis of a specific layer 
with respect to the optic axis of the previous layer is given by 
\begin{equation}
\label{def-q}
\Delta\xi=h\frac{\pi}{q}\,,\quad q\in\lec{2,3,4,\dots}\ric\,,
\end{equation}
 where $h = +1$ for structural right-handedness
and $h = -1$ for structural left-handedness. With the assumption that the constitutive parameters are independent of frequency
(which is an approximation of limited physical validity \cite{KK}), the Reusch pile shall display two families
of circular Bragg regimes. Circular Bragg regimes in the \textit{first} family are conjectured to exist
with center wavelengths
\begin{equation}
\label{fam1}
\lambdao^{1;p,q}= \frac{\lambdao^{1;0,q}}{pq+1}\,,\qquad p\in\lec{0,1,2,3,\dots}\ric\,, \quad  q\in\lec{2,3,4,\dots}\ric\,,
\end{equation}
whereas circular Bragg regimes   in the \textit{second} family are predicted
with center wavelengths
\begin{equation}
\label{fam2}
\lambdao^{2;p,q}= \frac{\lambdao^{1;0,q}}{pq+q-1}\,,\qquad p\in\lec{0,1,2,3,\dots}\ric\,, \quad  q\in\lec{2,3,4,\dots}\ric\,,
\end{equation}
where $\lambdao^{1;0,q}$ depends on the constitutive parameters and period of the Reusch pile and
normal incidence has been assumed.  
Equations~(\ref{fam1}) and (\ref{fam2}) will require modification for
frequency-dependent constitutive parameters \cite{Jianwei,Moyer}. Furthermore,
there may be other circular Bragg
regimes that are not captured by the conjectures (\ref{fam1}) and (\ref{fam2}).

When $h=1$ and the Reusch pile has a sufficiently large number of
periods, the reflectance of an incident right-circularly polarized (RCP)
 plane wave is very high but that of an incident left-circularly polarized (LCP) plane wave is very low in the first family
 of circular Bragg regimes,
 whereas  the reflectance of an incident LCP
 plane wave is very high but that of an incident RCP plane wave is very low in the second family.
 When $h=-1$ and the Reusch pile has a sufficiently large number of
periods, the reflectance of an incident LCP
 plane wave is very high but that of an incident RCP plane wave is very low in the first family,
 whereas  the reflectance of an incident RCP
 plane wave is very high but that of an incident LCP plane wave is very low in the second family of circular Bragg regimes.

Since
$\lambdao^{1;p,2}=\lambdao^{2;p,2}\, \forall {p}$, 
the Reusch pile is classified as \textit{equichiral} for $q=2$. If $q \geq 3$, the Reusch pile is classified as
\textit{ambichiral}. As $q\to \infty$, the Reusch pile is  \textit{finely chiral}  with
reflectance and transmittance characteristics similar to that  chiral liquid crystals \cite{Nityananda,Garoff,Oldano,Abdulhalim} and chiral
sculptured thin films \cite{STFbook,Erten2015,McAtee2018}.
Equichiral Reusch piles exhibit the  Bragg phenomenon  without differentiating between
LCP and RCP plane waves \cite{ambi,Rudakova}, as has been experimentally verified \cite{Fiallo1}.

Given that $\lambdao^{1;p+1,q}<\lambdao^{1;p,q}$ and $\lambdao^{2;p+1,q}<\lambdao^{2;p,q}$, it would be arduous
to experimentally verify the multiplicity of circular Bragg regimes with a single Reusch pile if $\lambdao^{1;0,q}$ lies in the visible regime.
Still, the exhibition of two adjacent circular Bragg regimes, the ones for $p=0$  from both families with $q$ fixed, has been experimentally confirmed with
some deviation from Eq.~(\ref{fam2}) due to frequency dependence of the constitutive parameters \cite{ambi,vanPopta}.

\subsection{Architected morphology}

With fixed $P$ but increasing $q$, Reusch piles offer an evolutionary perspective on the roles that architected
morphology can play in diverse optical phenomena. Such studies have been
undertaken for transmission-mode optical activity \cite{Joly3,Joly4,Babaei1}, circular Bragg phenomenon \cite{vanPopta,Abdulhalim2008,Babaei2},
and surface-wave propagation \cite{FL2014}. In this chapter, I deploy Reusch piles to study the evolution of the geometric phases
of the reflected and transmitted plane waves when a Reusch pile is illuminated by a plane wave.

The 2004 classification \cite{ambi} of Reusch piles excludes (i) columnar thin films \cite{HWbook,STFbook} and (ii) chevronic sculptured thin films \cite{VML},
because these materials with architected morphology do not exhibit the circular Bragg phenomenon. But their inclusion is necessary to understand the relationship,
or lack thereof, between the geometric phase and morphology. Therefore, I augmented Eq.~(\ref{def-q}) to
\begin{equation}
\label{def-Nlyr}
\Delta\xi=h\frac{2\pi}{\Nlyr}\,,\quad \Nlyr \in\lec{1,2,3,4,\dots}\ric\,.
\end{equation}
Whereas $\Nlyr=1$  for columnar thin films and $\Nlyr = 2$  for chevronic thin films, $\Nlyr =4$ 
for equichiral Reusch piles and $\Nlyr\in\lec{6,8,10,\dots}\ric$ for ambichiral Reusch piles in the 2004
classification \cite{ambi}. In addition, $\Nlyr\in\lec3,5,7,\dots\ric$,
which were  not explicitly included in the 2004 classification, are also possible while maintaining a full turn of $2\pi$ of the optic axes
within the thickness $P$. 

The conjectures (\ref{fam1}) and (\ref{fam2}) still apply with $q$ replaced by $\Nlyr/2$,
 $\Nlyr \in\lec{2,3,4,\dots}\ric$. The case of $\Nlyr=1$ has to be excluded, because a columnar thin film
 lacks periodicity. Equation (\ref{fam1}) indicates that
 \begin{equation}
 \label{seq1}
 \frac{\lambdao^{1;p,\Nlyr/2}}{  \lambdao^{1;0,\Nlyr/2}} \in
 \left\{
 \begin{array}{l}
\lec1,\frac{1}{2},\frac{1}{3},\frac{1}{4},\dots\ric\,,
\\[5pt]
\lec1,\frac{2}{5},\frac{2}{8},\frac{2}{11},\dots\ric\,,
\\[5pt]
\lec1,\frac{1}{3},\frac{1}{5},\frac{1}{7},\dots\ric\,,
\\[5pt]
\lec1,\frac{2}{7},\frac{2}{12},\frac{2}{17},\dots\ric\,,
\\[5pt]
\lec1,\frac{1}{4},\frac{1}{7},\frac{1}{10},\dots\ric\,,
\end{array}
\right.
\qquad
\Nlyr=
\left\{
\begin{array}{l}
2\,\\[5pt]
3\,\\[5pt]
4\,\\[5pt]
5\, \\[5pt]
6\,
\end{array}
\right.\,,
\end{equation}
and
 \begin{equation}
 \label{seq2}
 \frac{\lambdao^{2;p,\Nlyr/2}}{  \lambdao^{1;0,\Nlyr/2}} \in
 \left\{
 \begin{array}{l}
\lec\infty,1,\frac{1}{2},\frac{1}{3},\dots\ric\,,
\\[5pt]
\lec2,\frac{2}{4},\frac{2}{7},\frac{2}{10},\dots\ric\,,
\\[5pt]
\lec1,\frac{1}{3},\frac{1}{5},\frac{1}{7},\dots\ric\,,
\\[5pt]
\lec \frac{2}{3},\frac{2}{8},\frac{2}{13},\frac{2}{18},\dots\ric\,,
\\[5pt]
\lec\frac{1}{2},\frac{1}{5},\frac{1}{8},\frac{1}{11},\dots\ric\,,
\end{array}
\right.
\qquad
\Nlyr=
\left\{
\begin{array}{l}
2\,\\[5pt]
3\,\\[5pt]
4\,\\[5pt]
5\, \\[5pt]
6\,
\end{array}
\right.\,,
\end{equation}
Thus, the largest finite value of $\lambdao^{\ell;p,\Nlyr/2}$, $\ell\in\lec1,2\ric$, is $\lambdao^{1;0,\Nlyr/2}$
with one exception: $\lambdao^{2;0,3/2}= 2 \lambdao^{1;0,3/2}$. Note that
$\lambdao^{1;p,1}=\lambdao^{2;p+1,1}$ for chevronic thin films and $\lambdao^{1;p,2}=\lambdao^{2;p,2}$
for equichiral Reusch piles, both instances of the two families of circular Bragg regimes not being distinct from each other
in their center wavelengths.
The sequences in conjectures (\ref{seq1}) and (\ref{seq2}) will change when the  frequency dependence of
the constitutive parameters cannot be ignored.

 This chapter is organized as follows. Section~\ref{sec:theory} provides 
the theoretical framework to calculate  the geometric phase of the reflected/transmitted plane wave
in relation to the incident plane wave. Numerical results are presented
and discussed in
Sec.~\ref{sec:nrd}, and the paper ends with key conclusions in Sec.~\ref{sec:cr}.
An $ \exp(-i \omega t)$  dependence on time $t$ is implicit, where $ \omega$ as the angular frequency.
With $ \epso$ and $\muo$, respectively,
denoting the permittivity and permeability  of free space,
the free-space wavenumber is denoted by $\ko 
= \omega \sqrt{\epso \muo}$,  and $\lambdao=2\pi/\ko$ is the free-space
wavelength. The Cartesian coordinate system $(x,y,z)$ is adopted.
Vectors are in boldface
 and unit vectors are additionally decorated by a caret on top. Dyadics are double underlined.
Column vectors are underlined and enclosed in square brackets. Matrixes are double underlined and enclosed in square brackets.

 \section{Theory} \label{sec:theory}
 
 \subsection{Relative permittivity dyadic}
The Reusch pile is taken to occupy the region $0<z<NP$, where $N\in\lec1,2,3,\dots\ric$ is the number
of periods, and the  relative permittivity dyadic
\begin{equation}
\=\eps_{\rm rel}(z+P)= \=\eps_{\rm rel}(z)\,,\quad 0 < z < (N-1)P\,
\end{equation}
is therefore periodic. 

The reference unit cell $0<z<P$ is subdivided into $\Nlyr$ layers.
The $n$-th layer, $n\in\lec1,2,\dots,\Nlyr\ric$, is delimited by the planes $z=z_{n-1}$ and $z=z_n$, 
where
\begin{equation}
z_m=m\frac{P}{\Nlyr}\,,\quad m\in\lec0,1,2,\dots,\Nlyr\ric\,.
\end{equation}
The relative permittivity dyadic in the $n$-th layer of the reference unit cell is given
by
\begin{eqnarray}
\nonumber
\=\eps_{\rm rel}(z) &=&\=Z(h;n,\Nlyr)\. \=Y(\chi)\.\left[\epsa\uz\uz+\epsb\ux\ux
+\epsc\uy\uy\right]\.\=Y^{-1}(\chi)\.\=Z^{-1}(h;n,\Nlyr)\,,
\\
&& z_{n-1}<z<z_n\,,\quad n\in\lec1,2,\dots,\Nlyr\ric\,.
\label{def-epsrel}
\end{eqnarray}
 The frequency-dependent relative permittivity scalars
$\epsa$, $\epsb$, and $\epsc$ capture   the orthorhombicity \cite{STFbook} of each layer.
 The tilt dyadic
\begin{equation}
\=Y(\chi) = \uy\uy + \left(\ux\ux+\uz\uz\right)\cos\chi
 +\left(\uz\ux-\ux\uz\right)\sin\chi
 \label{SY-def}
 \end{equation}
contains $\chi\in[ {0,\pi/2}]$ as an angle of inclination with respect to the $xy$ plane.
The structural handedness is captured
by the rotation dyadic
\begin{eqnarray}
\nonumber
&&
\=Z(h;n,\Nlyr)=\uz\uz +\left(\ux\ux+\uy\uy\right)\cos\les{ (n-1)h\frac{2\pi}{\Nlyr}}\ris
+\left(\uy\ux-\ux\uy\right)\sin\les{ (n-1)h\frac{2\pi}{\Nlyr}}\ris\,,
\\
&&
\qquad
n\in\lec1,2,\dots,\Nlyr\ric\,.
\label{Sz-def}
\end{eqnarray} 

Examination of Eq.~(\ref{def-epsrel}) reveals the absence of periodicity only for $\Nlyr=1$.
However, the structural period is not always $P$ for all values of $\chi$. The structural period is $P$
for $\chi\in(0,\pi/2)$, but $P/2$ for $\chi\in\lec0,\pi/2\ric$. This distinction has been partially noted earlier \cite{LVaeu}
for continuously chiral materials (i.e., in the limit $\Nlyr\to\infty$).

Examination of Eq.~(\ref{def-epsrel})   reveals also the absence of structural handedness  for $\Nlyr\in\lec1,2\ric$.
Additionally, structural handedness is  absent for $\Nlyr=4$ provided that $\chi\in\lec0,\pi/2\ric$.

\subsection{Boundary-Value Problem}
The half-space $z < 0$ is the region of incidence and reflection, while
the half-space $z > L$ is the region of transmission.
A plane
wave, propagating in the half--space
$z \leq 0$ at an angle $\thetainc\in[0,\pi/2)$ with respect to the $z$ axis and at an angle $\psi\in[0,2\pi)$ with respect
to the $x$ axis in the $xy$ plane, is incident on the Reusch pile.
The electric  field phasor associated
with the incident plane wave is represented as \cite{STFbook}
\begin{subequations}
\begin{eqnarray}
\Ei&=& \les \frac{\le i\sp - \pinc \ri}{\sqrt{2}} \, \aL -\, \frac{\le i\sp +\pinc
\ri}{\sqrt{2}} \, \aR \ris
\exp{\les i\kappa \le x\cos\psi + y\sin\psi \ri
\ris} \exp{\left(i\ko z{\cos\thetainc}\right)}
\label{eq15a} \\
&=& \le \as \sp + \ap\pinc\ri  \exp{\les i\kappa \le x\cos\psi + y\sin\psi \ri
\ris} \exp{\left(i\ko z{\cos\thetainc}\right)}\,,
\label{eq15b}
\qquad   z < 0 \, ,
\end{eqnarray}
\end{subequations}
where
 \begin{equation}
 \left.\begin{array}{l}
\kappa =
\ko\sin{\thetainc}\\[5pt]
\sp=-\ux\sin\psi + \uy \cos\psi \\[5pt]
\#p_\pm=\mp\le \ux \cos\psi + \uy \sin\psi \ri \cos{\thetainc} 
 + \uz \sin{\thetainc}
\end{array}
\right\}
\, .
\end{equation}
The amplitudes of the LCP  and  the 
 RCP components of the incident plane wave, denoted by 
$\aL$ and
$\aR$, respectively, in Eq.~(\ref{eq15a}) are assumed to be known. Alternatively,
$\as$ and $\ap$ are the known amplitudes of the perpendicular- and parallel-polarized
components, respectively, in Eq.~(\ref{eq15b}).

The   electric   field phasor of the reflected plane wave is expressed as
\begin{subequations}
\begin{eqnarray}
\Er&=&-\, \les \frac{\le i\sp - \pref \ri}{\sqrt{2}} \, {\rL} -\, \frac{\le i\sp +
\pref \ri}{\sqrt{2}} \, {\rR} \ris  \exp{\les i\kappa \le x\cos{\psi} + y\sin{\psi} \ri
\ris} \exp\left({-i\ko   z\cos{\thetainc}}\right) 
\\
&=& \le \rs\sp+\rp\pref\ri \exp{\les i\kappa \le x\cos{\psi} + y\sin{\psi} \ri
\ris} \exp\left({-i\ko   z\cos{\thetainc}}\right) \, ,
 \qquad z < 0 \, ,
\end{eqnarray}
\end{subequations}
and the electric   field phasor of the transmitted plane wave is represented as
\begin{subequations}
\begin{eqnarray}
\Et&=& \les \frac{\le i\sp - \pinc \ri}{\sqrt{2}} \, {\tL} -\, \frac{\le i\sp +\pinc
\ri}{\sqrt{2}} \, {\tR} \ris  \exp{\les i\kappa \le x\cos{\psi} + y\sin{\psi} \ri
\ris} \exp\les{i\ko   (z-NP)\cos{\thetainc}}\ris \, , 
\\
&=& \le \ts\sp +\tp\pinc\ri    \exp{\les i\kappa \le x\cos{\psi} + y\sin{\psi} \ri
\ris}  
   \exp\les{i\ko   (z-NP)\cos{\thetainc}}\ris \, , 
\qquad z > NP \, .
\end{eqnarray}
\end{subequations}

The reflection amplitudes ${\rL}$ and ${\rR}$ as well as the transmission
amplitudes ${\tL}$ and ${\tR}$ (equivalently, $\rs$, $\rp$, $\ts$, and $\tp$)
are unknown and require the solution
of a boundary-value problem. The most straightforward
technique requires the use of  the 4$\times$4 transfer-matrix method, whose
details are available elsewhere
\cite{MLbook,STFbook}.
Thereafter, the total  reflectance
\begin{subequations}
\begin{equation}
\label{def-R}
R=\frac{\vert\rL\vert^2+\vert\rR\vert^2}{\vert\aL\vert^2+\vert\aR\vert^2} =\frac{\vert\rs\vert^2+\vert\rp\vert^2}{\vert\as\vert^2+\vert\ap\vert^2} 
\end{equation}
and the total
transmittance
\begin{equation}
\label{def-T}
T=\frac{\vert\tL\vert^2+\vert\tR\vert^2}{\vert\aL\vert^2+\vert\aR\vert^2} =\frac{\vert\ts\vert^2+\vert\tp\vert^2}{\vert\as\vert^2+\vert\ap\vert^2} 
\end{equation}
\end{subequations}
can be calculated.

\subsection{Poincar\'e spinors}

The Stokes parameters of the incident plane wave are given by \cite{Jackson}
\begin{equation}
\label{Stokes-tra}
\left.\begin{array}{l}
\sinczero = \vert\aL\vert^2+\vert\aR\vert^2 =\vert\as\vert^2+\vert\ap\vert^2
\\[5pt]
\sincone =2\,{\rm Re}\left(\aL\,\aR^\ast\right)=\vert\ap\vert^2-\vert\as\vert^2
\\[5pt]
\sinctwo =2\,{\rm Im}\left(\aL\,\aR^\ast\right)=2\,{\rm Re}\left(\as\,\ap^\ast\right)
\\[5pt]
\sincthree = \vert\aR\vert^2-\vert\aL\vert^2=2\,{\rm Im}\left(\as\,\ap^\ast\right)
\end{array}
\right\}\,,
\end{equation} 
where $^\ast$ denotes the complex conjugate.
The angles $\alphainc$
and $\betainc$ can be  calculated using the foregoing equations in Eqs.~(\ref{def-alphabeta}),
followed by the Poincar\'e spinor $\bphitinc$ from  Eq.~(\ref{def-PS}).

The Stokes parameters of the reflected plane wave are given by \cite{Jackson}
\begin{equation}
\label{Stokes-ref}
\left.\begin{array}{l}
\srefzero = \vert\rL\vert^2+\vert\rR\vert^2 =\vert\rs\vert^2+\vert\rp\vert^2
\\[5pt]
\srefone =2\,{\rm Re}\left(\rL\,\rR^\ast\right)=\vert\rp\vert^2-\vert\rs\vert^2
\\[5pt]
\sreftwo =2\,{\rm Im}\left(\rL\,\rR^\ast\right)=2\,{\rm Re}\left(\rs\,\rp^\ast\right)
\\[5pt]
\srefthree = \vert\rR\vert^2-\vert\rL\vert^2=2\,{\rm Im}\left(\rs\,\rp^\ast\right)
\end{array}
\right\}\,,
\end{equation} 
from which
the angles $\alpharef$ and $\betaref$ as well as the Poincar\'e spinor $\bphitref$ can be calculated using
Eqs.~(\ref{def-alphabeta}) and (\ref{def-PS}).
Calculation of the Stokes parameters of the transmitted plane wave, the angles $\alphatra$ and $\betatra$,
and the Poincar\'e spinor $\bphittra$ follows the same route.

Note that $\bPhirefR=\bPhitraR\equiv0$ in both Figs.~\ref{theta-Nlyr1}  and \ref{psi-Nlyr1}, because of the structure of $\bphitinc$ for
an incident RCP plane wave \cite{Lakh2024josab,Lakh2024pra}. The other six geometric phases $\bPhiref_{\ell}$ and $\bPhitra_{\ell}$,
$\ell\in\left\{s,p,L\right\}$,
are generally non-zero and
their spectral dependencies may not follow those of the corresponding total remittances defined in Eqs.~(\ref{def-R}) and (\ref{def-T}).

 \section{Numerical results} \label{sec:nrd}
 
 Calculations were made of the total reflectance $R_{\ell}$,  total transmittance $T_{\ell}$, 
 reflection-mode geometric phase $\bPhiref_{\ell}$, and transmission-mode geometric phase $\bPhitra_{\ell}$,
 $\ell\in\left\{s,p,R,L\right\}$.
The subscripts in these quantities indicate the polarization state of the incident plane wave:
 perpendicular ($s$), parallel ($p$), left-circular ($L$), or right-circular ($R$).
 
In order to incorporate causal frequency-dependent constitutive
 parameters \cite{Frisch,Silva}
in calculations, single-resonance Lorentzian functions were assumed 
for $\epsa$, $\epsb$, and $\epsc$
as follows \cite{Kittel}:
\begin{equation}
\label{resonance}
\eps_{\ell}(\lambdao) = 1+ \frac{p_{\ell}}
{1 + \le\frac{1}{N_{\ell}}  - i \frac{\lambda_{\ell}}{\lambdao} \ri^2}\,,\quad \ell\in\lec{a,b,c}\ric\,.
\end{equation}
The oscillator strengths are determined by the values of $p_{\ell}$,  $\lambda_{\ell} (1 + N_{\ell}^{-2})^{-1/2}  $  are the  resonance wavelengths, and $\lambda_{\ell}/N_{\ell}$ are the  resonance linewidths. 
 The parameters used for  the theoretical results reported here are as follows:  $p_{\rm a} = 2.3$, $p_{\rm b} =3.0$, $p_{\rm c} =2.2 $, $\lambda_{\rm a} = \lambda_{\rm c} =260$~nm, $\lambda_{\rm b} = 270$~nm,   and
$N_{\rm a} = N_{\rm b} =N_{\rm c}=130$. Furthermore, $\chi = 37\deg$, $N=15$,  and $P =  300$~nm were fixed. 
Calculations were made for primarily for $\lambdao\in[400,800]$~nm.

\subsection{Columnar thin film ($\Nlyr=1$)}

Figure~\ref{theta-Nlyr1} presents the spectrums of $R_{\ell}$,   $T_{\ell}$, $\bPhiref_{\ell}$, and   $\bPhitra_{\ell}$, $\ell\in\left\{s,p,R,L\right\}$,
for $\thetainc\in[0\deg,90\deg)$ and $\psi=0\deg$.  Figure~\ref{psi-Nlyr1} presents the analogous spectrums for $\thetainc=0\deg$ and $\psi\in[0\deg,360\deg)$.
As the Reusch pile reduces to a single columnar thin film of thickness $NP$ when  $\Nlyr = 1$,  no Bragg phenomenon can be
exhibited, the columnar thin film being effectively a homogeneous biaxial-dielectric continuum \cite{HWbook,MLpnfa}
whose relative permittivity dyadic does not depend on $h$. 

In    Figs.~\ref{theta-Nlyr1}  and  \ref{psi-Nlyr1}, the total linear remittances (i.e., $R_{\rm s}$, $R_{\rm p}$, $T_{\rm s}$, and $T_{\rm p}$)  depend on the polarization state of the incident linearly polarized
plane wave, but the total circular remittances (i.e., $R_{\rm R}$, $R_{\rm L}$, $T_{\rm R}$, and $T_{\rm L}$) do not depend on the polarization state of the incident
circularly polarized plane wave. The plots of all eight
quantities   show Fabry--Perot resonances \cite{FP}, as expected from a homogeneous dielectric slab. The spectral dependences
of  $\bPhiref_{\ell}$ do not follow those of $R_\ell$, and the spectral dependences of
 $\bPhitra_{\ell}$ do not follow those of $T_\ell$,
$\ell\in\left\{s,p,L\right\}$.

\subsection{Chevronic thin film ($\Nlyr=2$)}

\begin{figure}[ht]
	\begin{center}
			\includegraphics[width=0.5\linewidth]{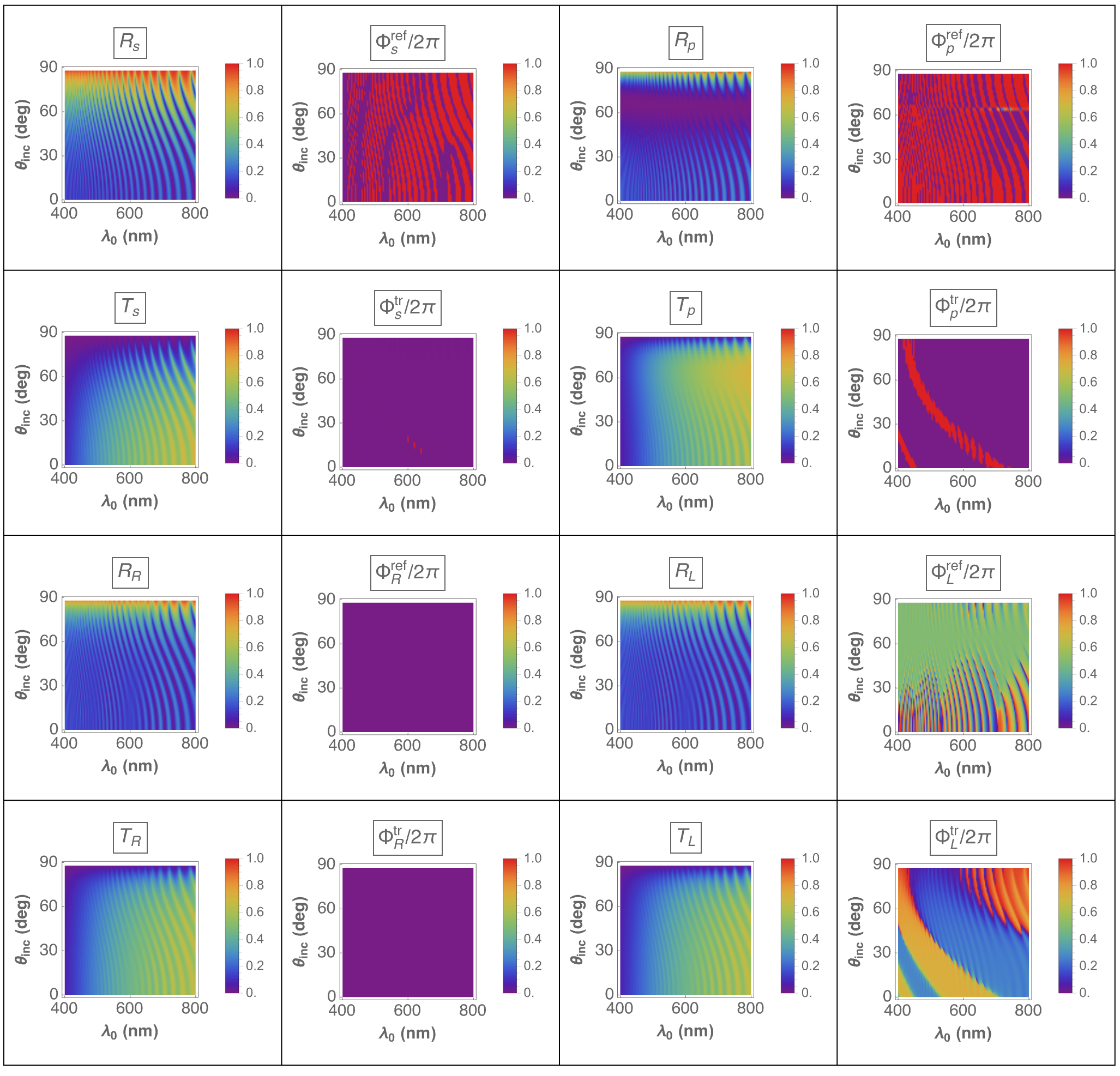}
	\end{center}
	\caption{Spectrums of $R_{\ell}$,   $T_{\ell}$, $\bPhiref_{\ell}$, and   $\bPhitra_{\ell}$, $\ell\in\left\{s,p,R,L\right\}$,
	for $\thetainc\in[0\deg,90\deg)$,
	when $\Nlyr=1$, $h=\pm1$, $N=15$,  $P=300$~nm, and $\psi=0\deg$.}
	\label{theta-Nlyr1}
\end{figure}

\begin{figure}[ht]
	\begin{center}
			\includegraphics[width=0.5\linewidth]{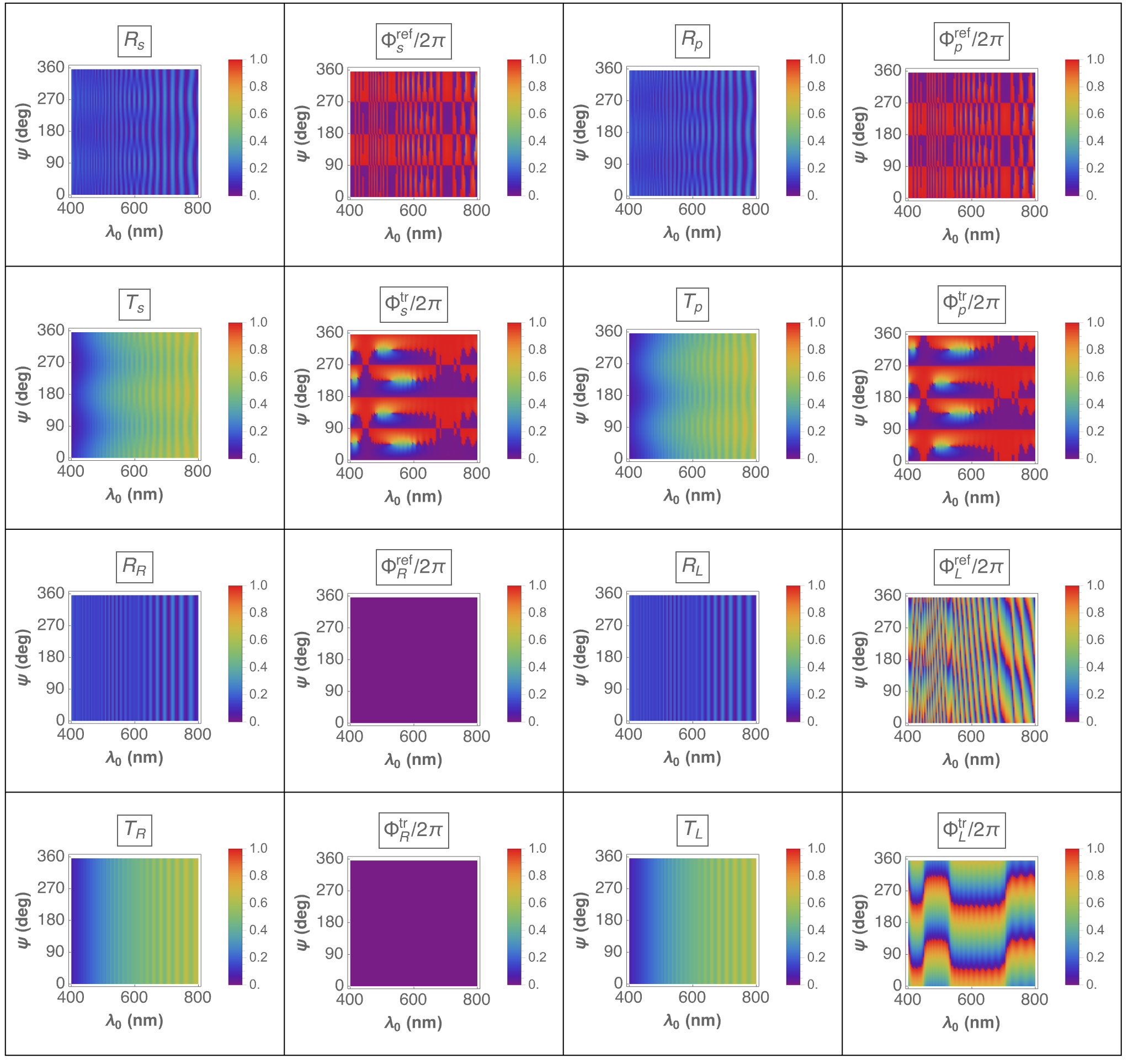}
	\end{center}
	\caption{Spectrums of $R_{\ell}$,   $T_{\ell}$, $\bPhiref_{\ell}$, and   $\bPhitra_{\ell}$, $\ell\in\left\{s,p,R,L\right\}$,
	for $\psi\in[0\deg,360\deg)$,
	when $\Nlyr=1$, $h=\pm1$, $N=15$,  $P=300$~nm, and $\thetainc=0\deg$.}
	\label{psi-Nlyr1}
\end{figure}

\begin{figure}[ht]
	\begin{center}
			\includegraphics[width=0.5\linewidth]{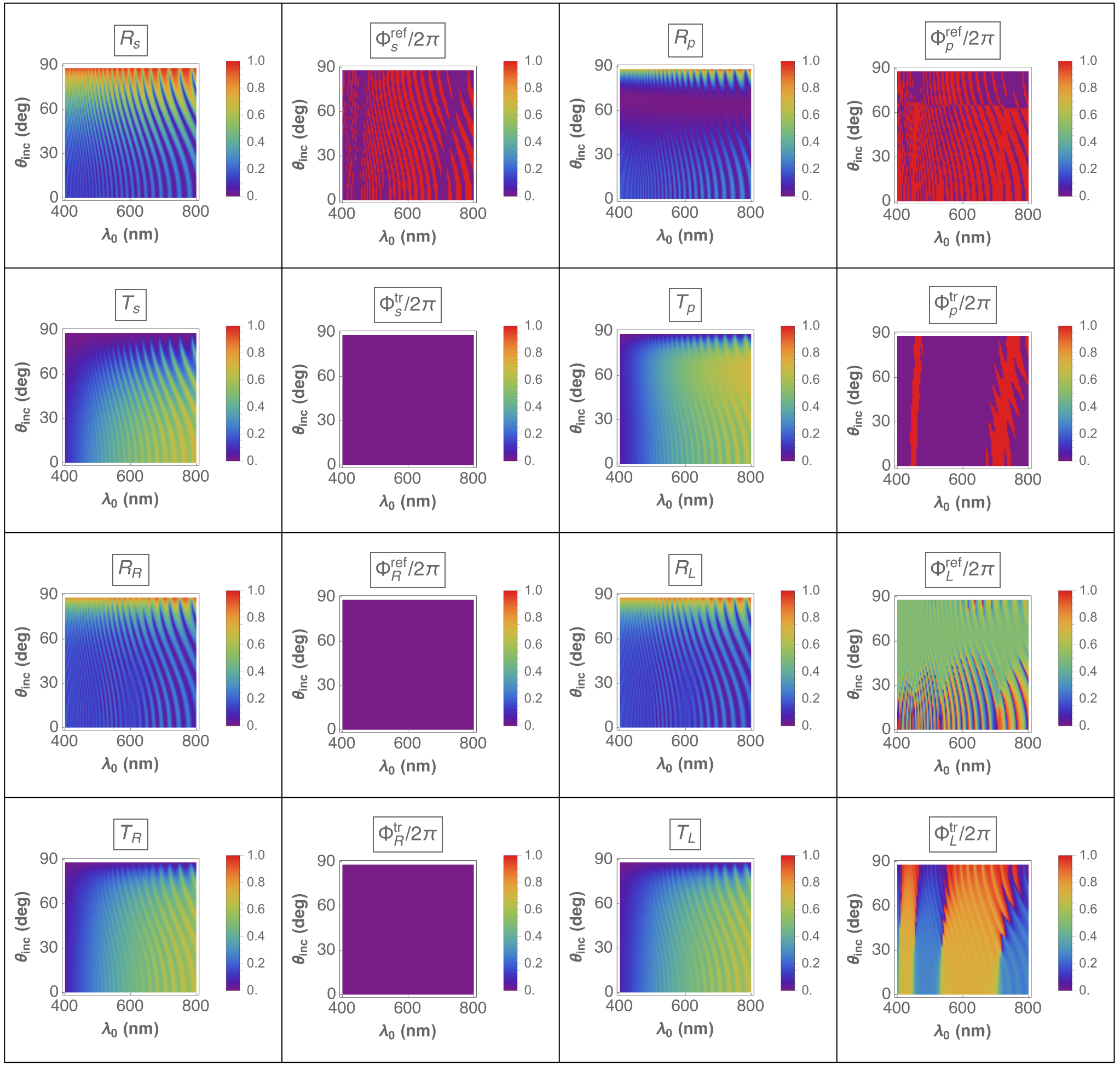}
	\end{center}
	\caption{Spectrums of $R_{\ell}$,   $T_{\ell}$, $\bPhiref_{\ell}$, and   $\bPhitra_{\ell}$, $\ell\in\left\{s,p,R,L\right\}$,
	for $\thetainc\in[0\deg,90\deg)$,
	when $\Nlyr=2$, $h=\pm1$, $N=15$,  $P=300$~nm, and $\psi=0\deg$.}
	\label{theta-Nlyr2}
\end{figure}

\begin{figure}[ht]
	\begin{center}
			\includegraphics[width=0.5\linewidth]{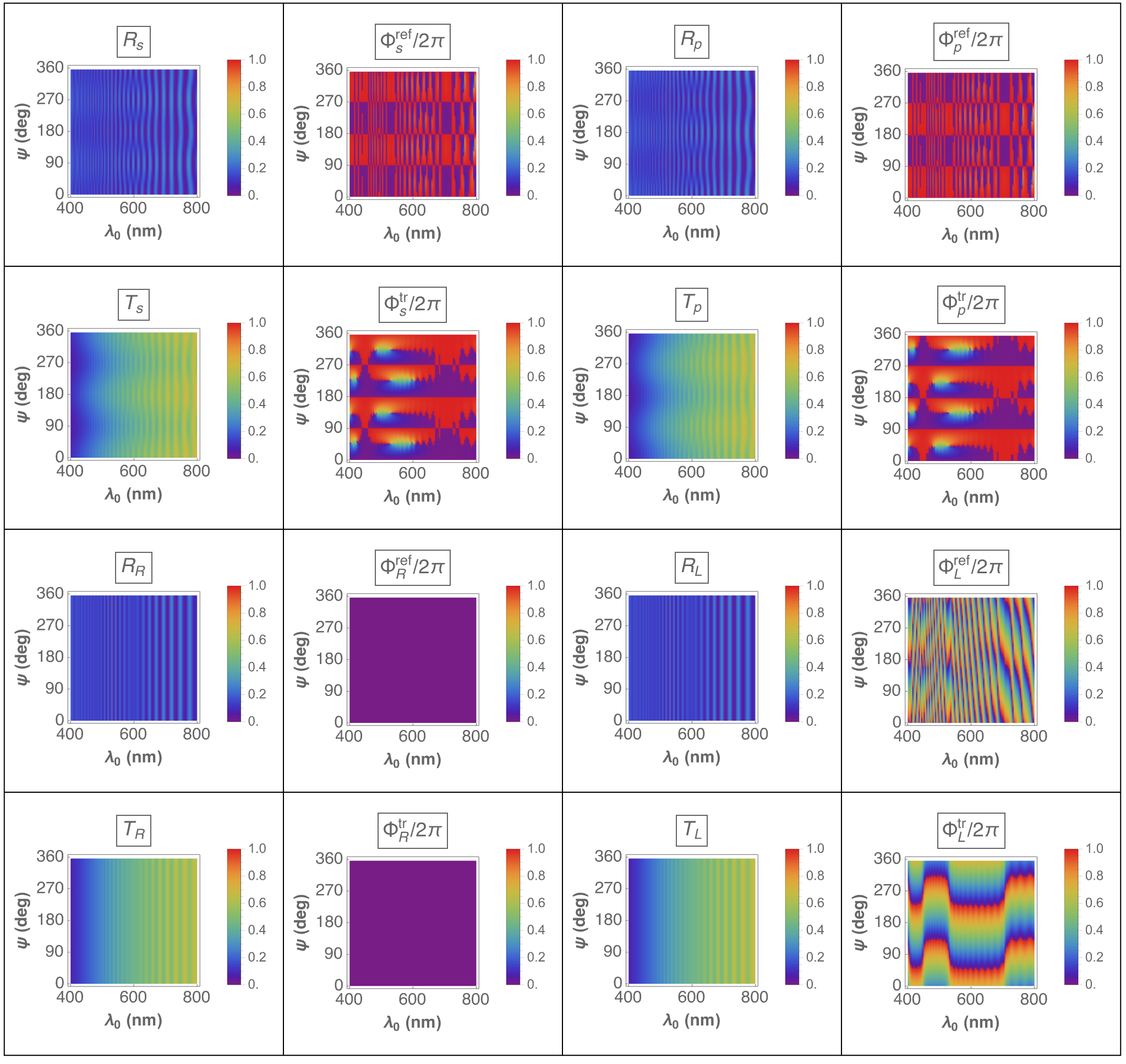}
	\end{center}
	\caption{Spectrums of $R_{\ell}$,   $T_{\ell}$, $\bPhiref_{\ell}$, and   $\bPhitra_{\ell}$, $\ell\in\left\{s,p,R,L\right\}$,
	for $\psi\in[0\deg,360\deg)$,
	when $\Nlyr=2$, $h=\pm1$, $N=15$,  $P=300$~nm, and $\thetainc=0\deg$.}
	\label{psi-Nlyr2}
\end{figure}

Figure~\ref{theta-Nlyr2} presents the spectrums of $R_{\ell}$,   $T_{\ell}$, $\bPhiref_{\ell}$, and   $\bPhitra_{\ell}$, $\ell\in\left\{s,p,R,L\right\}$,
for $\thetainc\in[0\deg,90\deg)$ and $\psi=0\deg$, and Fig.~\ref{psi-Nlyr2} for $\thetainc=0\deg$ and $\psi\in[0\deg,360\deg)$. 
Neither the total remittances nor the corresponding geometric phases  depend on whether $h=1$ or $h=-1$
  in these two figures, just as in Figs.~\ref{theta-Nlyr1} and  \ref{psi-Nlyr1}, because chevronic
  thin films lack structural handedness.

No trace of the Bragg phenomenon  is evident in the plots of the total remittances in Figs.~\ref{theta-Nlyr2} and  \ref{psi-Nlyr2}.
There is no doubt that
the Reusch pile is structurally periodic for $\Nlyr = 2$, but that structural periodicity does not translate into electromagnetic periodicity
for all incidence conditions. Indeed for normal incidence, the interface $z=P/2$ in the reference unit
cell is reflectionless \cite{LM1994,SM1998}.

Elsewhere, theoretical research has shown that the Bragg phenomenon is not exhibited by a
chevronic thin film for normal and near-normal
incidence, which conclusion has been validated experimentally \cite{VML}.
It is difficult to distinguish between the total remittance plots for
$\Nlyr=1$ (Figs.~\ref{theta-Nlyr1} and  \ref{psi-Nlyr1}) and $\Nlyr=2$  (Figs.~\ref{theta-Nlyr2} and  \ref{psi-Nlyr2}).
Theory also indicates
vestigial manifestation of the Bragg phenomenon is possible for highly oblique incidence \cite{VML}, but a clear
signature cannot be discerned in Fig.~\ref{theta-Nlyr2}.

The geometric phases $\bPhiref_{\ell}$ and $\bPhitra_{\ell}$,
$\ell\in\left\{s,p,L\right\}$, are not identically zero
in Figs.~\ref{theta-Nlyr2}  and \ref{psi-Nlyr2}. Furthermore, although the total remittance plots in Fig.~\ref{theta-Nlyr1}
are virtually indistinguishable from their counterparts in Fig.~\ref{theta-Nlyr2}, the dependences of $\bPhi^{\rm ref}_{\rm s,p}$,
$\bPhi^{\rm tr}_{\rm p}$, and $\bPhi^{\rm ref}_{\rm L}$ on $\thetainc$ in those two figures show clear differences. These differences
indicate that the geometric phases of the reflected and the transmitted plane waves are affected by morphology much more than the total
remittances, this observation having been previously made only for the 
geometric phase of the transmitted plane wave \cite{Lakh2024josab,Lakh2024pra}. Note, however, that
the plots of $\bPhi^{\rm ref}_{\rm s,p}$,
$\bPhi^{\rm tr}_{\rm p}$, and $\bPhi^{\rm ref}_{\rm L}$ in Figs.~\ref{psi-Nlyr1} and \ref{psi-Nlyr2} are identical, because
the interface $z=P/2$ in the middle of the reference unit cell of the chevronic thin film is electromagnetically inconsequential  for
normal incidence \cite{LM1994,SM1998}.

\subsection{Ambichiral Reusch pile ($\Nlyr=3$)}

\begin{figure}[ht]
	\begin{center}
			\includegraphics[width=0.5\linewidth]{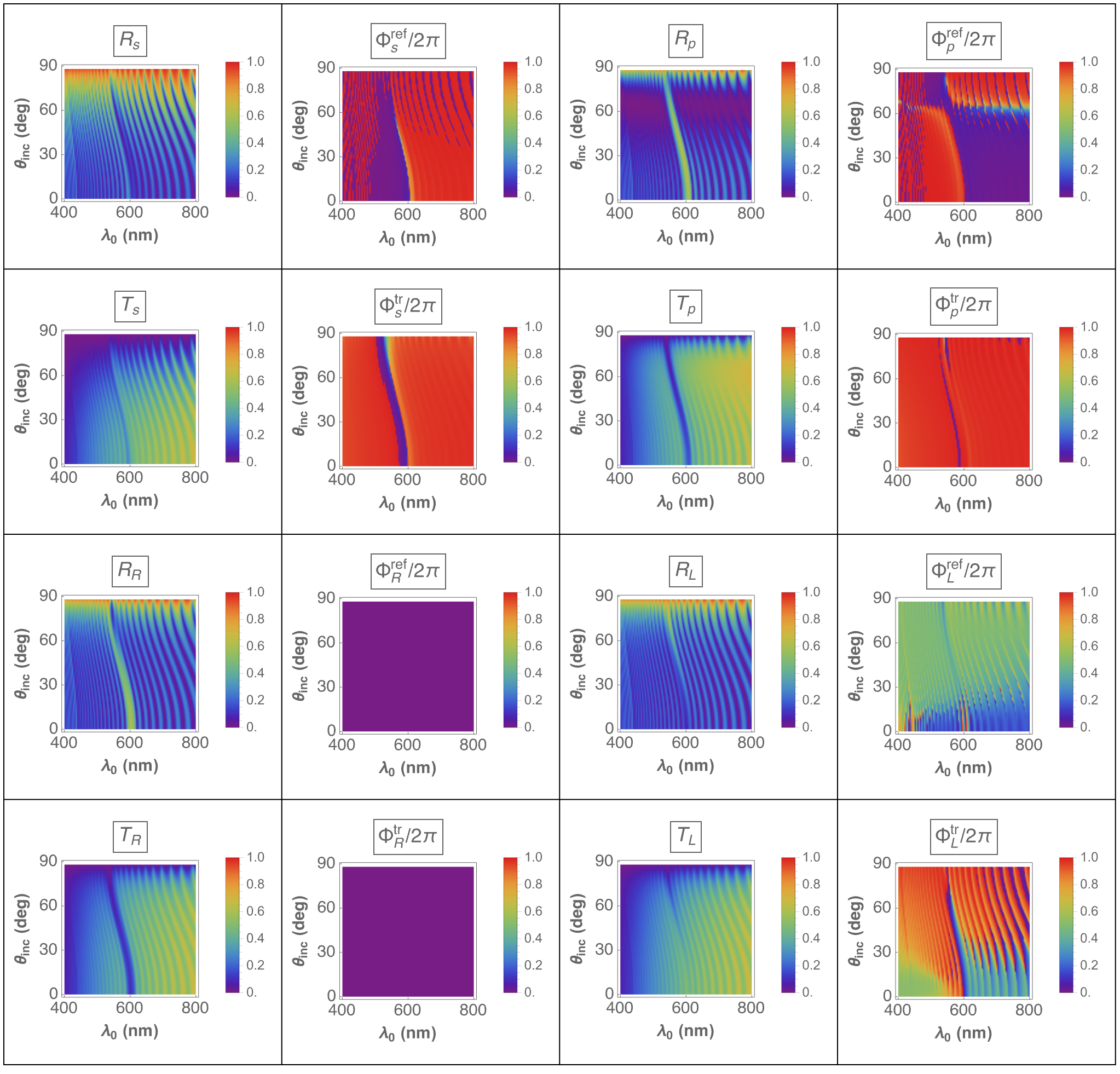}
	\end{center}
	\caption{Spectrums of $R_{\ell}$,   $T_{\ell}$, $\bPhiref_{\ell}$, and   $\bPhitra_{\ell}$, $\ell\in\left\{s,p,R,L\right\}$,
	for $\thetainc\in[0\deg,90\deg)$,
	when $\Nlyr=3$, $h= 1$, $N=15$,  $P=300$~nm, and $\psi=0\deg$.}
	\label{theta-Nlyr3-hpos}
\end{figure}

\begin{figure}[ht]
	\begin{center}
			\includegraphics[width=0.5\linewidth]{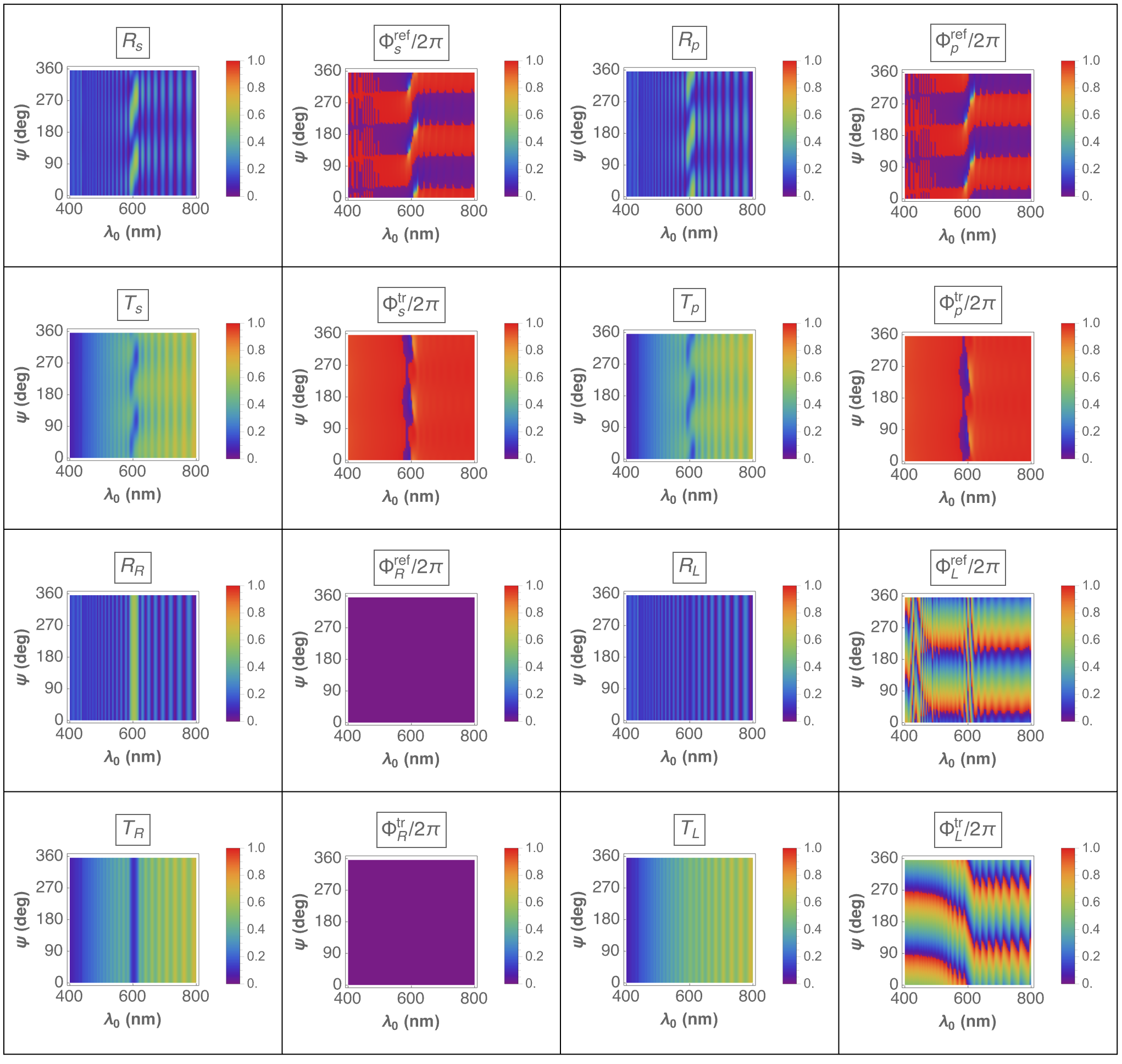}
	\end{center}
	\caption{Spectrums of $R_{\ell}$,   $T_{\ell}$, $\bPhiref_{\ell}$, and   $\bPhitra_{\ell}$, $\ell\in\left\{s,p,R,L\right\}$,
	for $\psi\in[0\deg,360\deg)$,
	when $\Nlyr=3$, $h= 1$, $N=15$,  $P=300$~nm, and $\thetainc=0\deg$.}
	\label{psi-Nlyr3-hpos}
\end{figure}

A Reusch pile with $\Nlyr=3$ possesses both structural handedness and structural periodicity. Although this Reusch
pile was not included in the 2004 classification \cite{ambi}, it should be considered
as ambichiral. Figure~\ref{theta-Nlyr3-hpos} presents the chosen
spectrums for $\thetainc\in[0\deg,90\deg)$ and $\psi=0\deg$, and Fig.~\ref{psi-Nlyr3-hpos} for $\thetainc=0\deg$ and $\psi\in[0\deg,360\deg)$,
when $h=1$. The  Bragg phenomenon is clearly evident as a deep blue trough in the plots of
$\TR$ and a corresponding ridge in the plots of $\RR$. This trough/ridge is centered at $\lambdao=602$~nm for $\thetainc=0\deg$ and
it  blueshifts with more oblique
incidence \cite{Erten2015}. The absence of analogous features in the plots of $\TL$ and $\RL$ supports the conclusion that this is
a \textit{circular Bragg phenomenon}. However, in comparison to a chiral sculptured thin film \cite{Erten2015}, the trough is present
in the plots of both $\TR$ and $\TL$, and the ridge in the plots of both $\RR$ and $\RL$, for highly oblique incidence. 

\begin{figure}[ht]
	\begin{center}
			\includegraphics[width=0.5\linewidth]{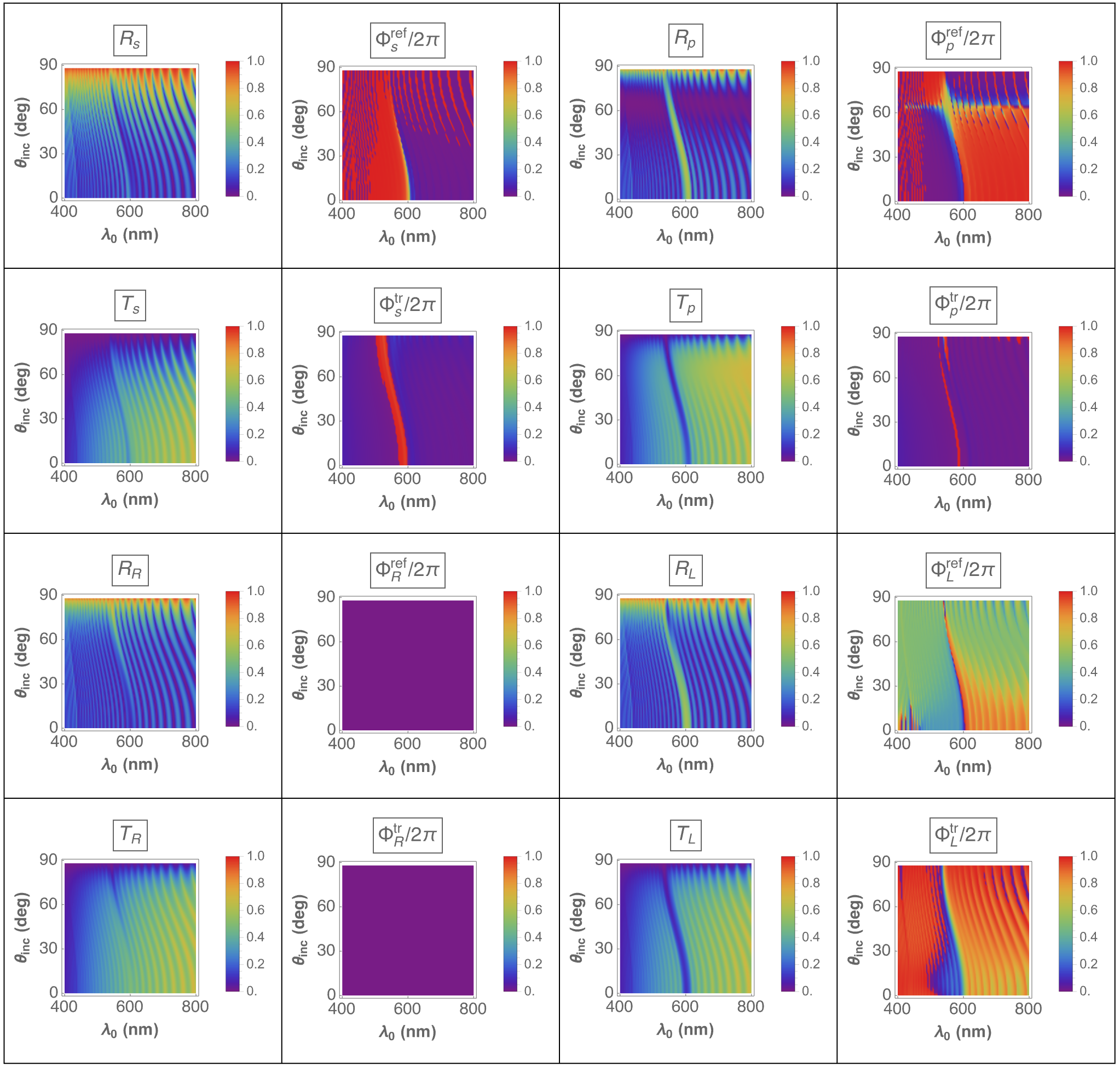}
	\end{center}
	\caption{Spectrums of $R_{\ell}$,   $T_{\ell}$, $\bPhiref_{\ell}$, and   $\bPhitra_{\ell}$, $\ell\in\left\{s,p,R,L\right\}$,
	for $\thetainc\in[0\deg,90\deg)$,
	when $\Nlyr=3$, $h=- 1$, $N=15$,  $P=300$~nm, and $\psi=0\deg$.}
	\label{theta-Nlyr3-hneg}
\end{figure}

\begin{figure}[ht]
	\begin{center}
			\includegraphics[width=0.5\linewidth]{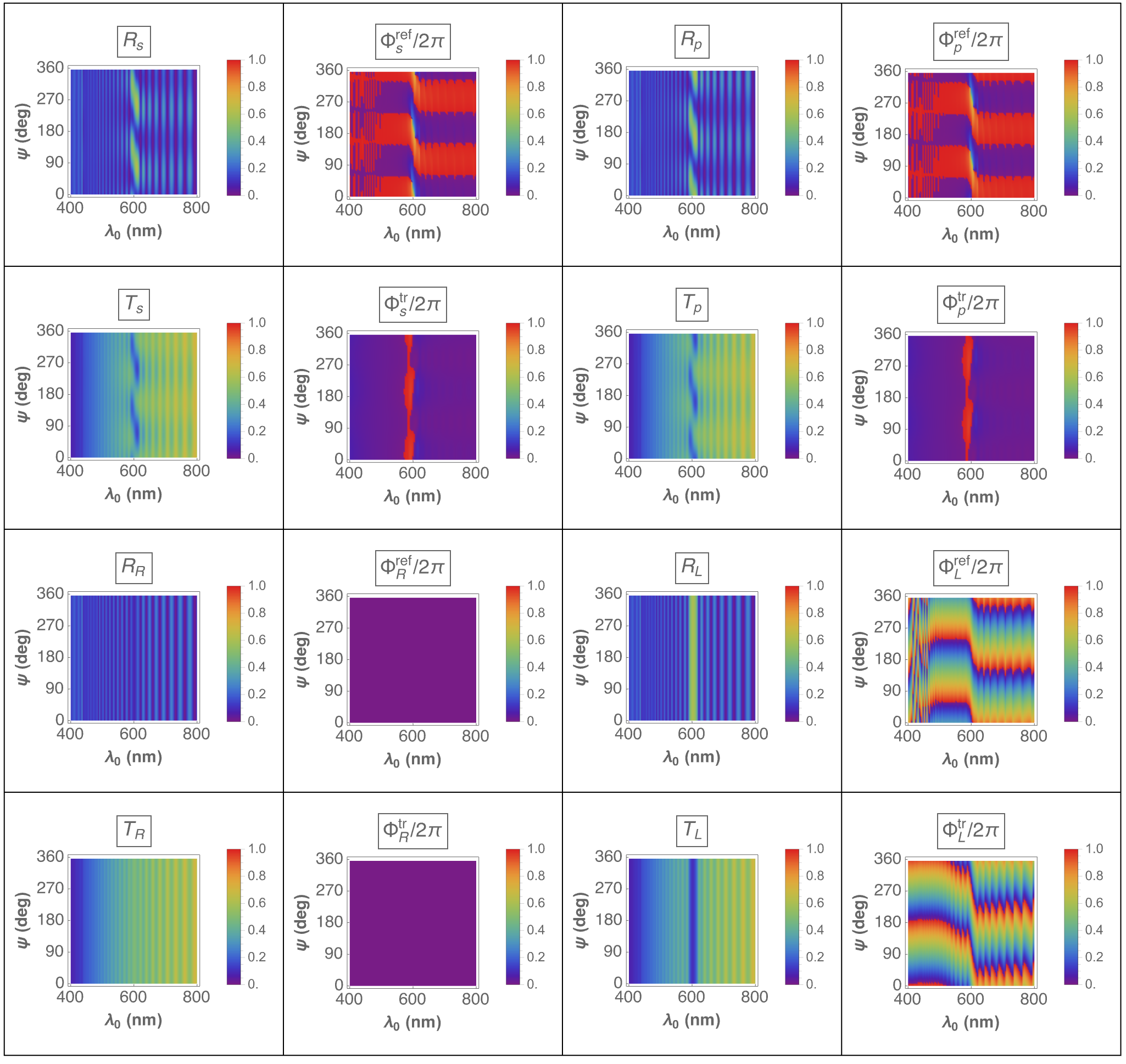}
	\end{center}
	\caption{Spectrums of $R_{\ell}$,   $T_{\ell}$, $\bPhiref_{\ell}$, and   $\bPhitra_{\ell}$, $\ell\in\left\{s,p,R,L\right\}$,
	for $\psi\in[0\deg,360\deg)$,
	when $\Nlyr=3$, $h= -1$, $N=15$,  $P=300$~nm, and $\thetainc=0\deg$.}
	\label{psi-Nlyr3-hneg}
\end{figure}

Analogous spectrums for $h=-1$ are presented in Figs.~\ref{theta-Nlyr3-hneg} and \ref{psi-Nlyr3-hneg}. The plots of $\lec\TR, \RR,\TL,\RL\ric$ for fixed
$\psi$
are interchanged with those of  $\lec\TL, \RL,\TR,\RR\ric$ for $2\pi-\psi$, in comparison to  Figs.~\ref{theta-Nlyr3-hpos} and \ref{psi-Nlyr3-hpos};
in other words,
\begin{equation}
\label{circ-symm}
\left.\begin{array}{ll}
\RR(h,\psi)=\RL(-h,2\pi-\psi)\,,\qquad&
\RL(h,\psi)=\RR(-h,2\pi-\psi)\\[5pt]
\TR(h,\psi)=\TL(-h,2\pi-\psi)\,,\qquad&
 \TL(h,\psi)=\TR(-h,2\pi-\psi)
\end{array}
\right\}\,.
\end{equation}

Since a linearly polarized plane wave can be decomposed into an RCP plane wave and an LCP plane wave, signatures of the circular
Bragg phenomenon are also to be found in the plots of linear remittances in Figs.~\ref{theta-Nlyr3-hpos}--\ref{psi-Nlyr3-hneg}. The following
symmetries are exhibited by the linear remittances:
\begin{equation}
\label{lin-symm}
\left.\begin{array}{ll}
\Rs(h,\psi)=\Rs(-h,2\pi-\psi)\,,\qquad&
\Rp(h,\psi)=\Rp(-h,2\pi-\psi)\\[5pt]
\Ts(h,\psi)=\Ts(-h,2\pi-\psi)\,,\qquad&
 \Tp(h,\psi)=\Tp(-h,2\pi-\psi)
\end{array}
\right\}\,.
\end{equation}

No symmetries are evident in the plots of geometric phases in Figs.~\ref{theta-Nlyr3-hpos}--\ref{psi-Nlyr3-hneg}, except that
$\bPhirefR=\bPhitraR\equiv0$ by virtue of the structure of the Poincar\'e spinor of a RCP plane wave \cite{Lakh2024josab,Lakh2024pra}.
The spectrums of $\bPhi_{\ell}^{\rm ref}$ and $\bPhi_{\ell}^{\rm tr}$, $\ell\in\lec s,p,L\ric$, are greatly affected when the structural handedness
is reversed.

\begin{figure}[ht]
	\begin{center}
			\includegraphics[width=0.4\linewidth]{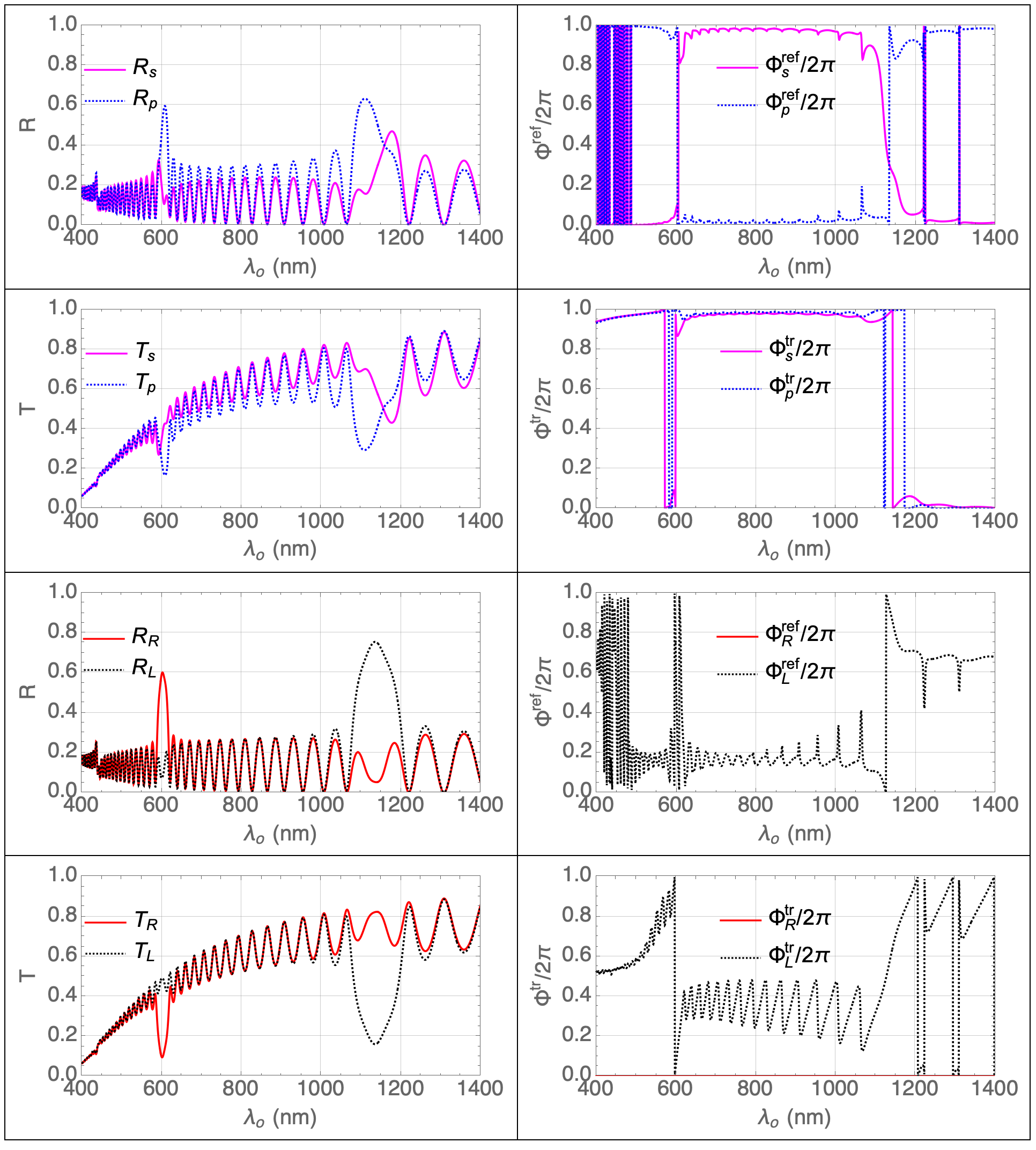}
	\end{center}
	\caption{Spectrums of $R_{\ell}$,   $T_{\ell}$, $\bPhiref_{\ell}$, and   $\bPhitra_{\ell}$, $\ell\in\left\{s,p,R,L\right\}$,
	for $\thetainc=0\deg$ and  $\psi=0\deg$,
	when $\Nlyr=3$, $h= 1$, $N=15$,  and $P=300$~nm.}
	\label{Nlyr3-hpos-extended}
\end{figure}

\begin{figure}[ht]
	\begin{center}
			\includegraphics[width=0.4\linewidth]{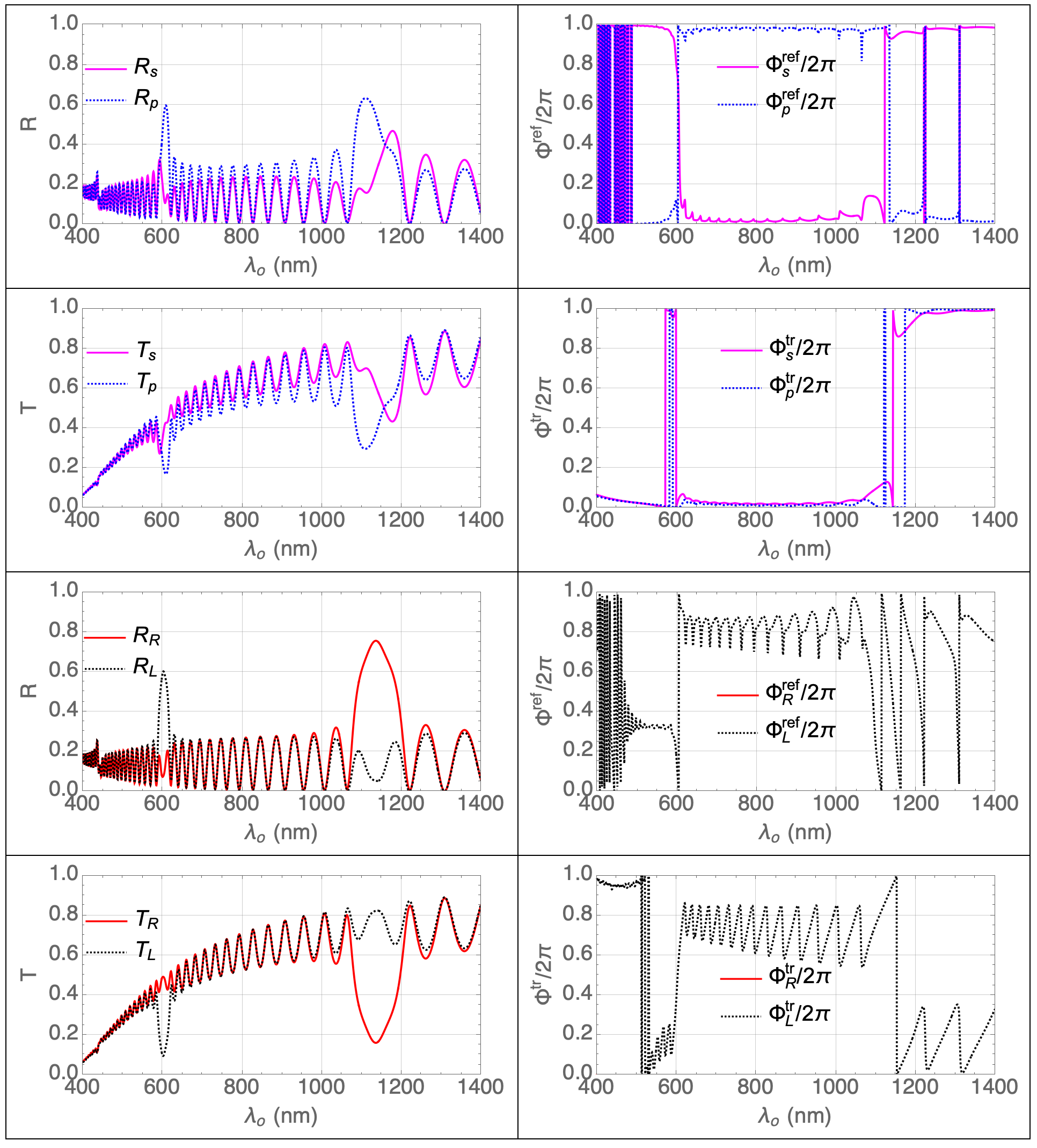}
	\end{center}
	\caption{Spectrums of $R_{\ell}$,   $T_{\ell}$, $\bPhiref_{\ell}$, and   $\bPhitra_{\ell}$, $\ell\in\left\{s,p,R,L\right\}$,
	for $\thetainc=0\deg$ and  $\psi=0\deg$,
	when $\Nlyr=3$, $h= -1$, $N=15$,  and $P=300$~nm.}
	\label{Nlyr3-hneg-extended}
\end{figure}

In order to confirm that $\Nlyr=3$ provides the exceptional case $\lambdao^{2;0,\Nlyr/2}>\lambdao^{1;0,\Nlyr/2}$,
calculations were also made for $\lambdao$ extending into the near-infrared
spectral regime. With $\thetainc=0\deg$ and $\psi=0\deg$ fixed, (i)
Fig.~\ref{Nlyr3-hpos-extended} presents the 
spectrums of $R_{\ell}$,   $T_{\ell}$, $\bPhiref_{\ell}$, and   $\bPhitra_{\ell}$, $\ell\in\left\{s,p,R,L\right\}$,
 for $h=1$, and Fig.~\ref{Nlyr3-hneg-extended} presents the same
spectrums  for $h=-1$. The remittance spectrums in both figures clearly show two circular Bragg regimes.
The first
is centered at $\lambdao^{1;0,3/2}=602$~nm, exactly as in Figs.~\ref{theta-Nlyr3-hpos}--\ref{psi-Nlyr3-hneg},
and it belongs to the first family described in Sec.~\ref{sec:RP}. The second
is centered at $\lambdao^{2;0,3/2}=1135$~nm and belongs to the second family described in Sec.~\ref{sec:RP}. 
Note that $\lambdao^{2;0,3/2}$ differs from $2\lambdao^{1;0,3/2}$ 
predicted by Eq.~(\ref{seq2}) because the relative permittivity scalars in
Eq.~(\ref{resonance}) are frequency dependent.  

Importantly, the spectral variations of $\bPhiref_{\ell}$ and   $\bPhitra_{\ell}$, $\ell\in\left\{s,p,L\right\}$,
are different in the two circular Bragg regimes in Figs.~\ref{Nlyr3-hpos-extended} and  \ref{Nlyr3-hneg-extended}.

\subsection{Equichiral Reusch pile ($\Nlyr=4$)}
The prefix \textit{equi}
in the classification \textit{equichiral} is justified for a Reusch pile with $\Nlyr=4$, since $\lambdao^{1;p,\Nlyr/2}= \lambdao^{2;p,\Nlyr/2}$~$\forall{p}$ when $\Nlyr=4$. The same Reusch pile is also structurally handed (with period $P$), so that  the suffix \textit{chiral}
in the classification \textit{equichiral} is also justified. However, there is a notable exception:
 the period equals $P/2$ and there is no structural handedness when $\chi\in\lec0,\pi/2\ric$.

\begin{figure}[ht]
	\begin{center}
			\includegraphics[width=0.5\linewidth]{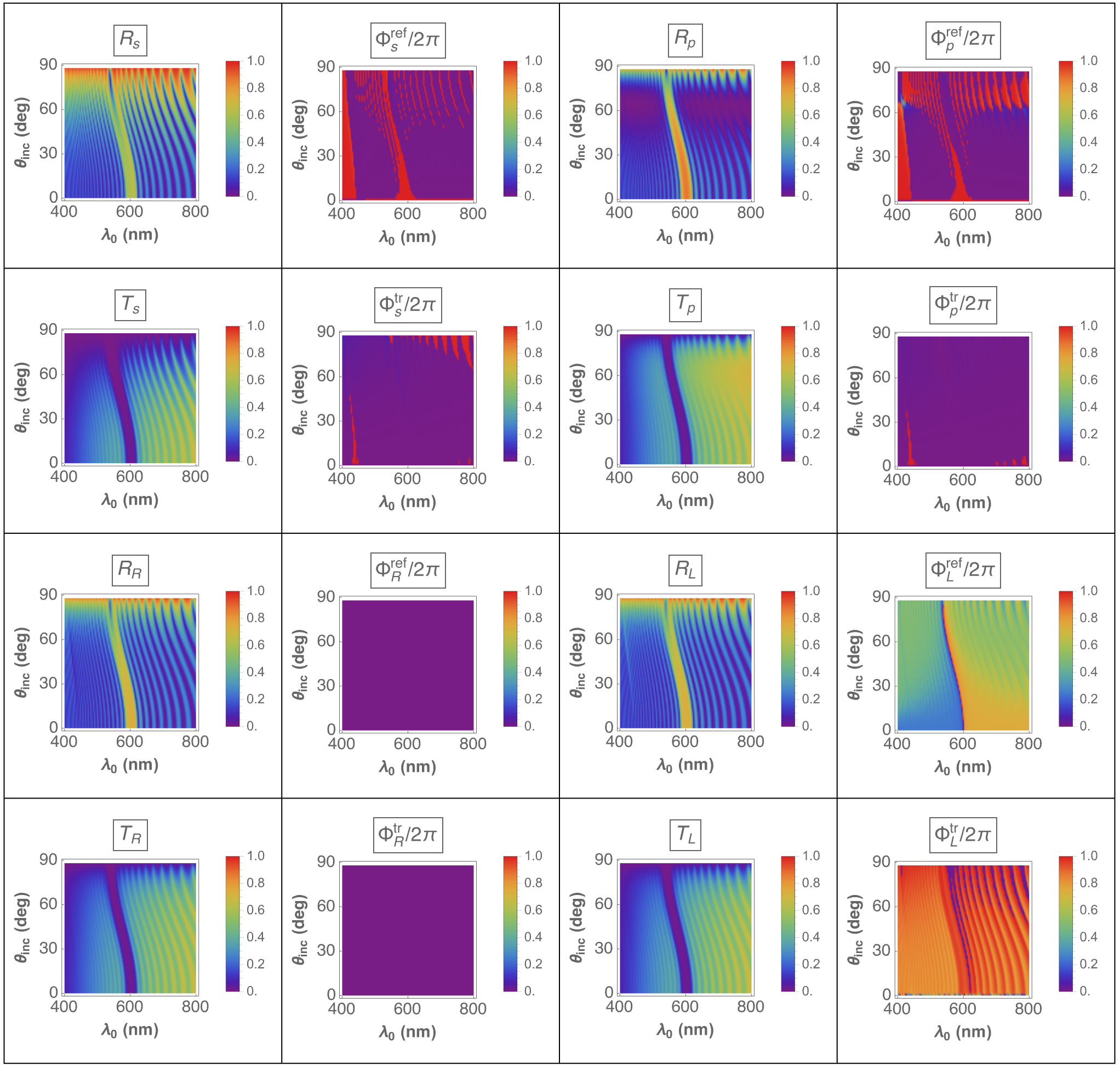}
	\end{center}
	\caption{Spectrums of $R_{\ell}$,   $T_{\ell}$, $\bPhiref_{\ell}$, and   $\bPhitra_{\ell}$, $\ell\in\left\{s,p,R,L\right\}$,
	for $\thetainc\in[0\deg,90\deg)$,
	when $\Nlyr=4$, $h= 1$, $N=15$,  $P=300$~nm, and $\psi=0\deg$.}
	\label{theta-Nlyr4-hpos}
\end{figure}

\begin{figure}[ht]
	\begin{center}
			\includegraphics[width=0.5\linewidth]{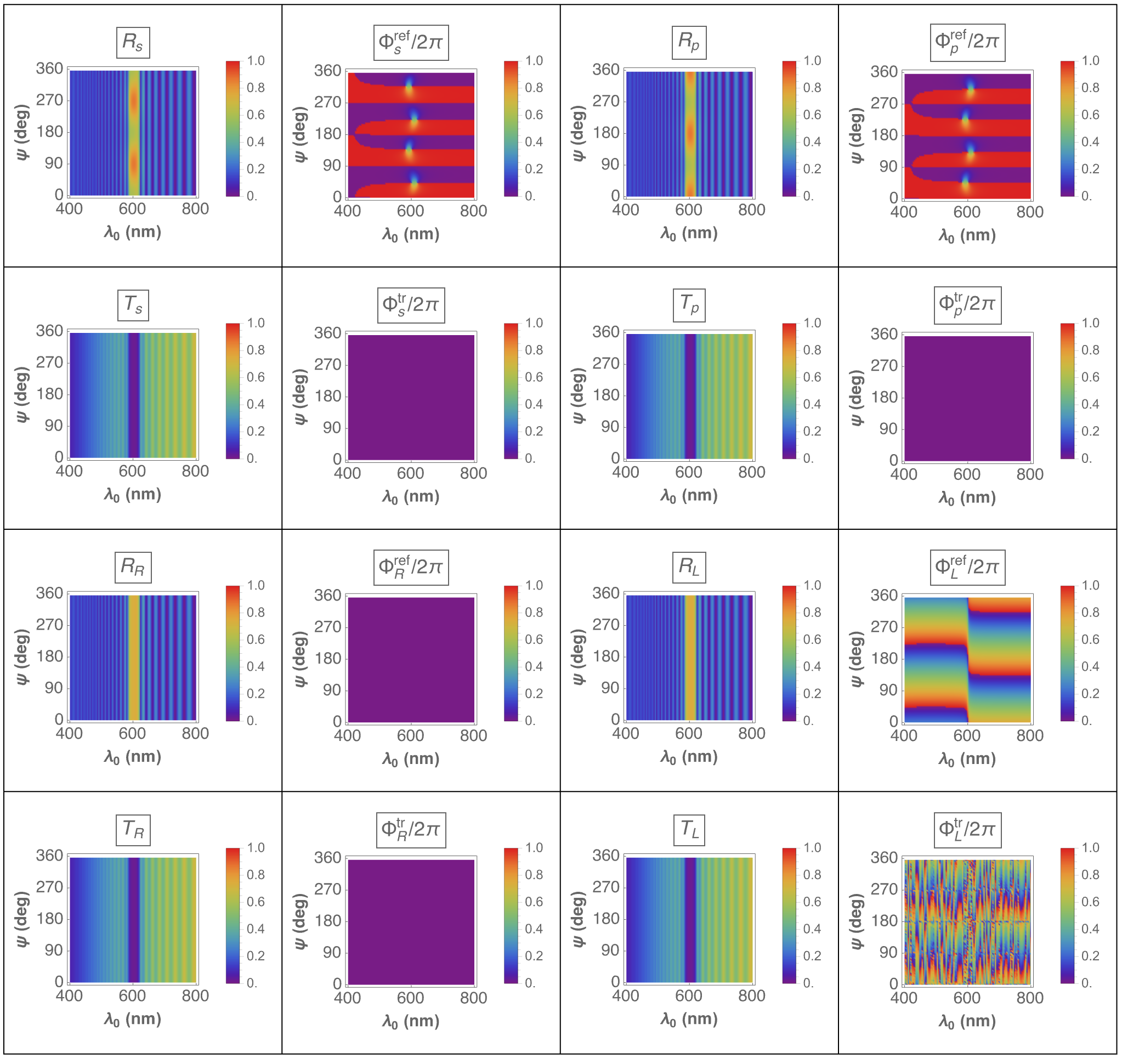}
	\end{center}
	\caption{Spectrums of $R_{\ell}$,   $T_{\ell}$, $\bPhiref_{\ell}$, and   $\bPhitra_{\ell}$, $\ell\in\left\{s,p,R,L\right\}$,
	for $\psi\in[0\deg,360\deg)$,
	when $\Nlyr=4$, $h= 1$, $N=15$,  $P=300$~nm, and $\thetainc=0\deg$.}
	\label{psi-Nlyr4-hpos}
\end{figure}

\begin{figure}[ht]
	\begin{center}
			\includegraphics[width=0.5\linewidth]{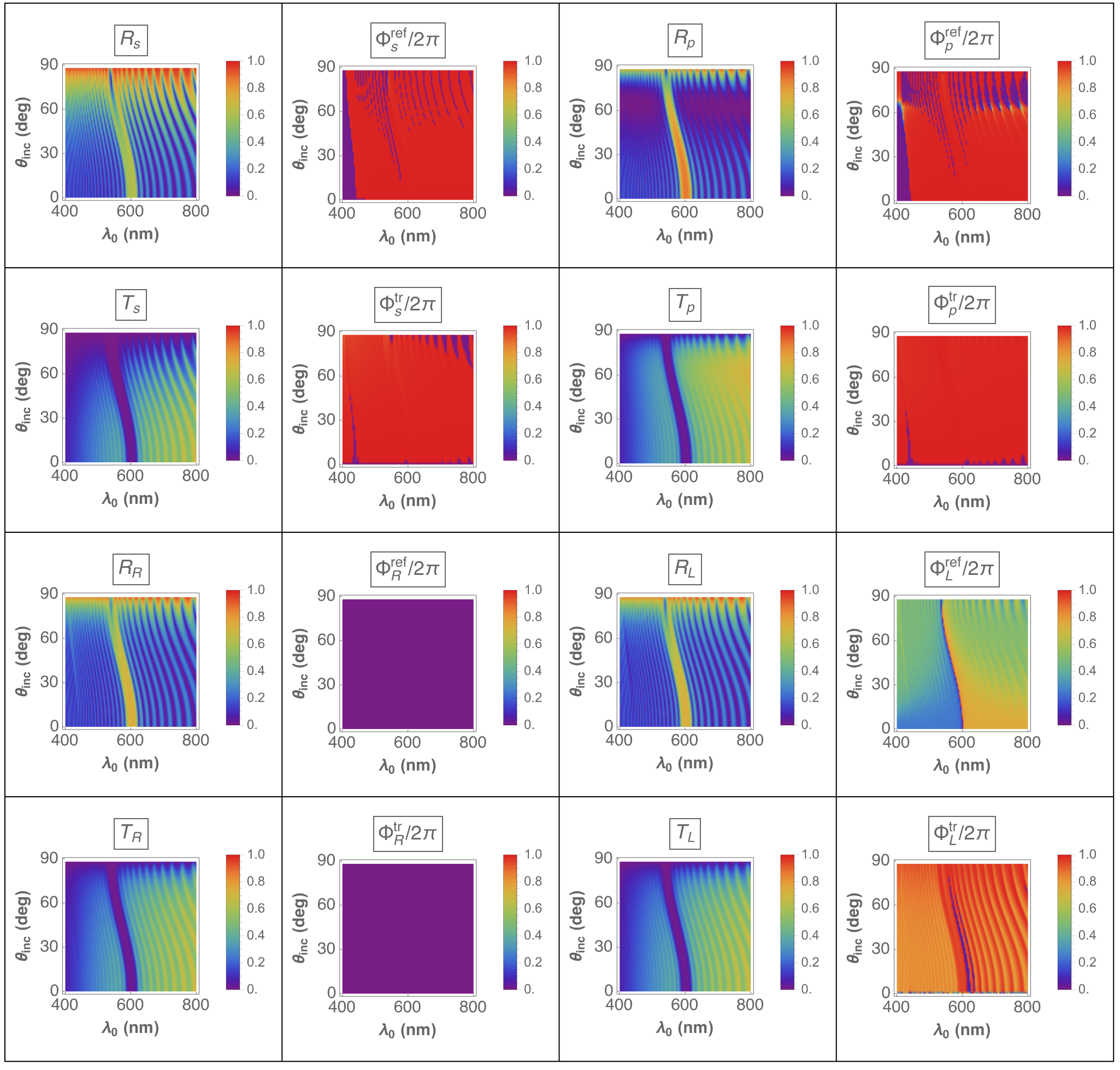}
	\end{center}
	\caption{Spectrums of $R_{\ell}$,   $T_{\ell}$, $\bPhiref_{\ell}$, and   $\bPhitra_{\ell}$, $\ell\in\left\{s,p,R,L\right\}$,
	for $\thetainc\in[0\deg,90\deg)$,
	when $\Nlyr=4$, $h= -1$, $N=15$,  $P=300$~nm, and $\psi=0\deg$.}
	\label{theta-Nlyr4-hneg}
\end{figure}

\begin{figure}[ht]
	\begin{center}
			\includegraphics[width=0.5\linewidth]{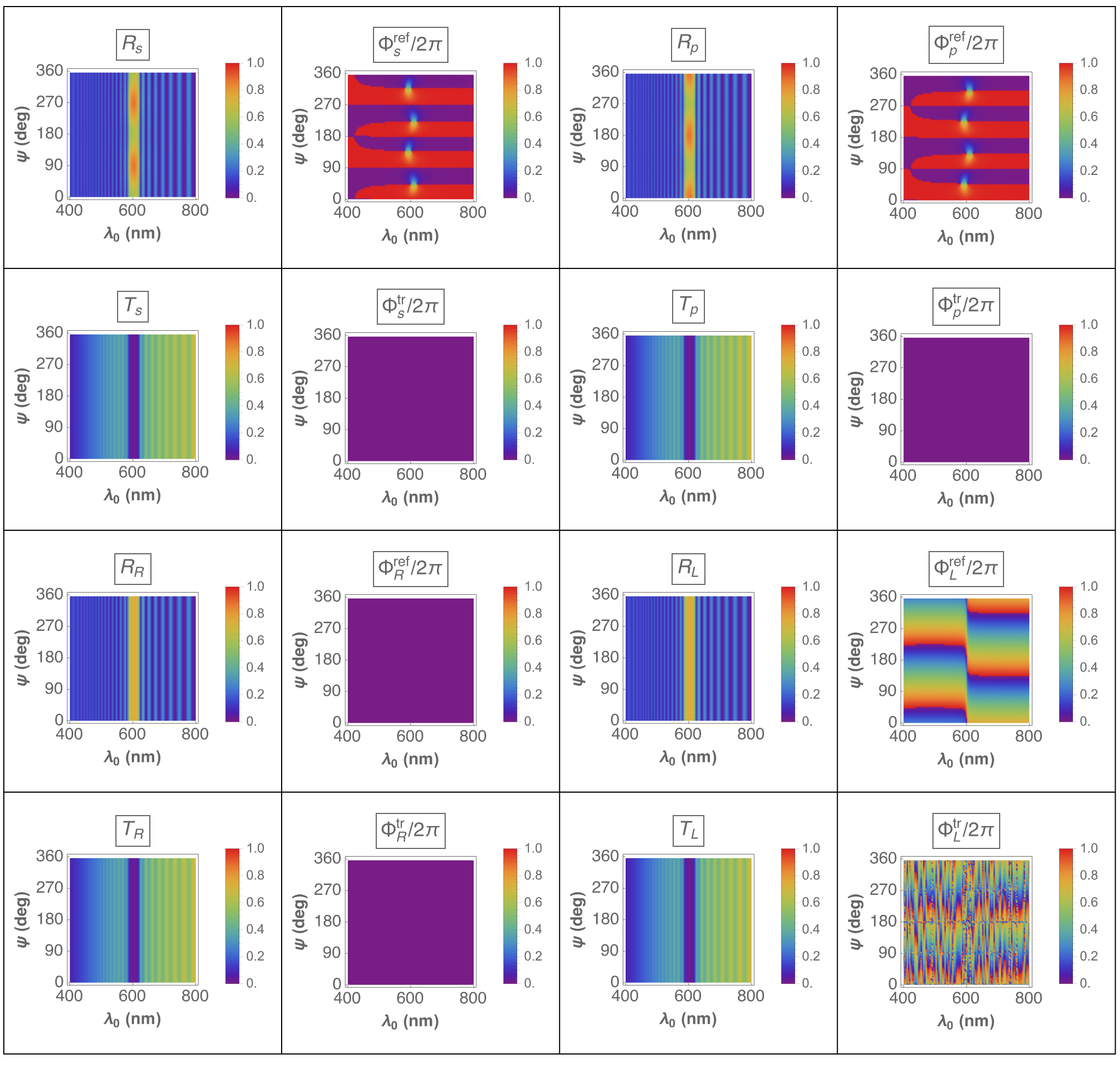}
	\end{center}
	\caption{Spectrums of $R_{\ell}$,   $T_{\ell}$, $\bPhiref_{\ell}$, and   $\bPhitra_{\ell}$, $\ell\in\left\{s,p,R,L\right\}$,
	for $\psi\in[0\deg,360\deg)$,
	when $\Nlyr=4$, $h= -1$, $N=15$,  $P=300$~nm, and $\thetainc=0\deg$.}
	\label{psi-Nlyr4-hneg}
\end{figure}

Figure~\ref{theta-Nlyr4-hpos} presents the chosen
spectrums for $\thetainc\in[0\deg,90\deg)$ and $\psi=0\deg$, and Fig.~\ref{psi-Nlyr4-hpos} for $\thetainc=0\deg$ and $\psi\in[0\deg,360\deg)$,
when $h=1$. Figures~\ref{theta-Nlyr4-hneg} and \ref{psi-Nlyr4-hneg} are the counterparts of those
two figures for $h=-1$. Calculations show that Eqs.~(\ref{circ-symm})
and (\ref{lin-symm}) hold for $\Nlyr=4$.

The  Bragg phenomenon is clearly evident as a deep blue trough in the plots of both linear and both circular transmittances,
and as  a corresponding ridge in the plots of both linear and both circular reflectances, regardless of the value of
$h\in\lec-1,1\ric$. These features are centered about $\lambdao=600$~nm for $\thetainc=0\deg$ and blueshift with more oblique
incidence. This Reusch pile   exhibits a polarization-universal bandgap that can be tuned
by adjusting the angle of incidence $\thetainc$, as has been verified experimentally \cite{Fiallo1}.

The plots of $\bPhi_{\ell}^{\rm ref}$ and $\bPhi_{\ell}^{\rm tr}$, $\ell\in\lec s,p,L\ric$, in general, contain clear evidence of the
 polarization-universal bandgap. However, whereas $\bPhi_{\ell}^{\rm ref}$ and $\bPhi_{\ell}^{\rm tr}$, $\ell\in\lec s,p\ric$, are greatly affected when the structural handedness
is reversed,  $\bPhi_{L}^{\rm ref}$ and $\bPhi_{L}^{\rm tr}$ are affected very little by the same reversal.

\subsection{Ambichiral and finely chiral Reusch piles ($\Nlyr\geq5$)}

The optical response characteristics for $\Nlyr\geq5$ are similar to those for $\Nlyr=3$, except that Eq. (\ref{seq2})
predicts
$\lambdao^{2;0,\Nlyr/2}/\lambdao^{1;0,\Nlyr/2}>1$ for $\Nlyr=3$ but $\lambdao^{2;0,\Nlyr/2}/\lambdao^{1;0,\Nlyr/2}<1$
for $\Nlyr\geq5$.

\begin{figure}[ht]
	\begin{center}
			\includegraphics[width=0.5\linewidth]{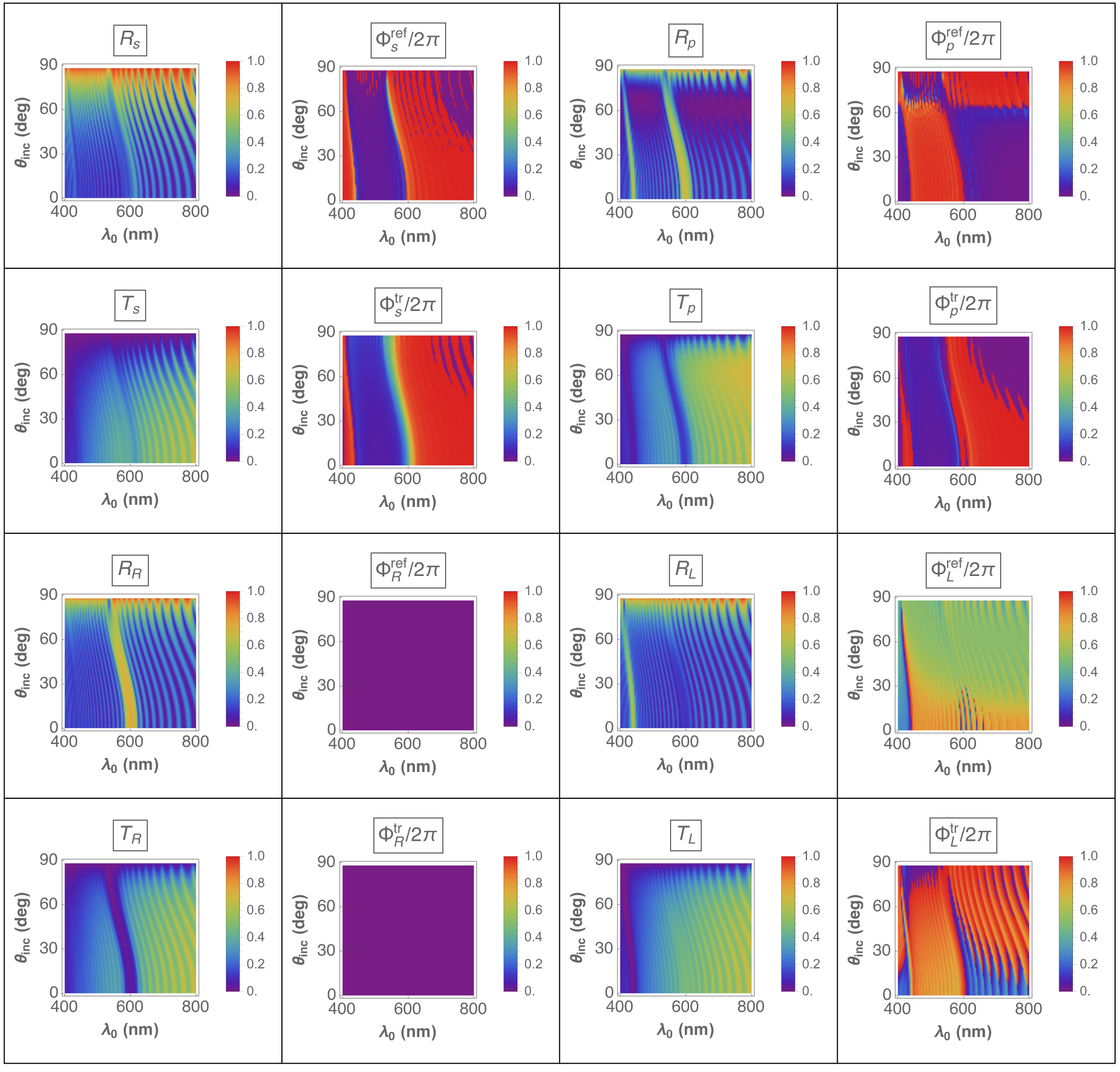}
	\end{center}
	\caption{Spectrums of $R_{\ell}$,   $T_{\ell}$, $\bPhiref_{\ell}$, and   $\bPhitra_{\ell}$, $\ell\in\left\{s,p,R,L\right\}$,
	for $\thetainc\in[0\deg,90\deg)$,
	when $\Nlyr=5$, $h= 1$, $N=15$,  $P=300$~nm, and $\psi=0\deg$.}
	\label{theta-Nlyr5-hpos}
\end{figure}

\begin{figure}[ht]
	\begin{center}
			\includegraphics[width=0.5\linewidth]{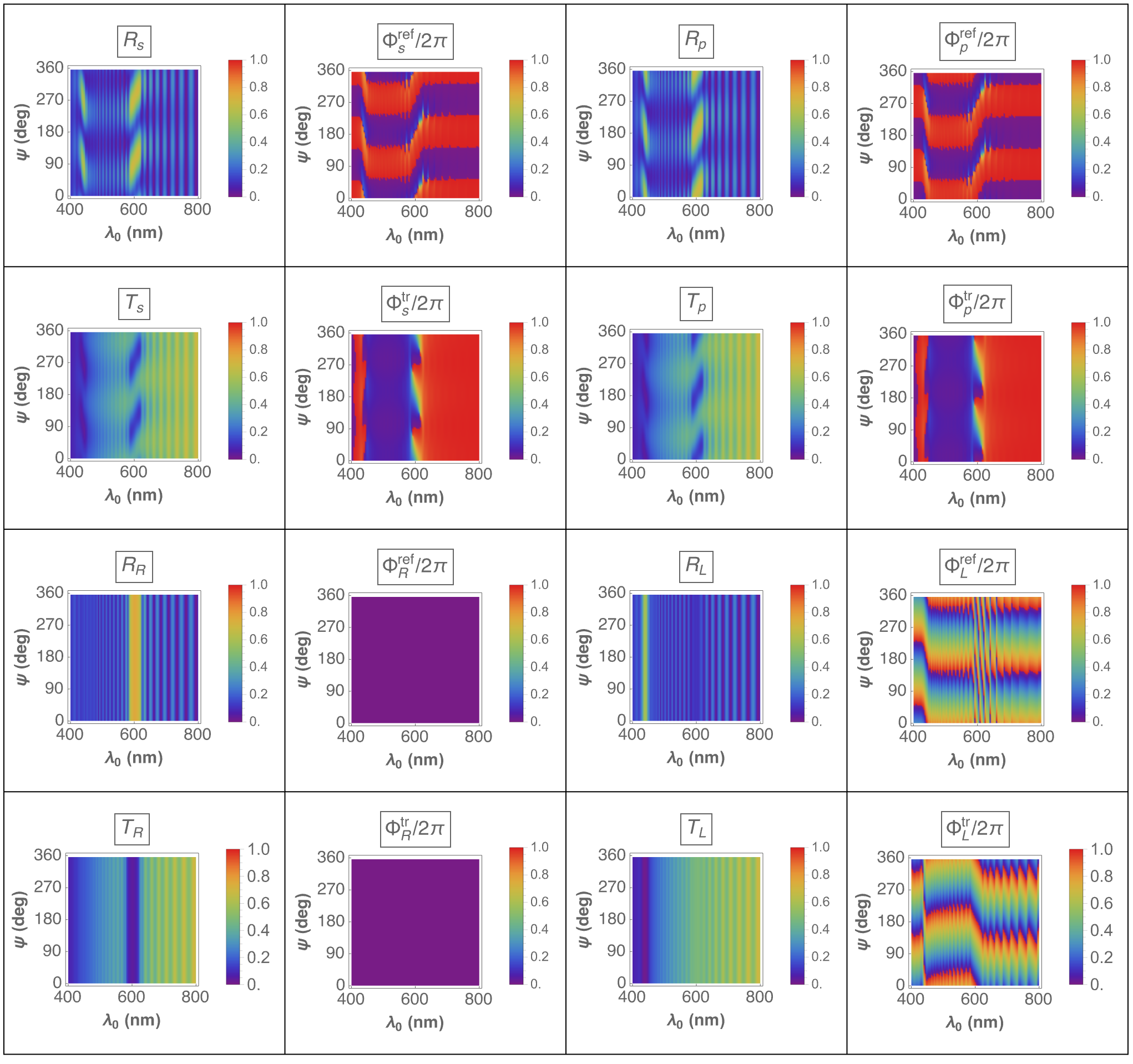}
	\end{center}
	\caption{Spectrums of $R_{\ell}$,   $T_{\ell}$, $\bPhiref_{\ell}$, and   $\bPhitra_{\ell}$, $\ell\in\left\{s,p,R,L\right\}$,
	for $\psi\in[0\deg,360\deg)$,
	when $\Nlyr=5$, $h= 1$, $N=15$,  $P=300$~nm, and $\thetainc=0\deg$.}
	\label{psi-Nlyr5-hpos}
\end{figure}

\begin{figure}[ht]
	\begin{center}
			\includegraphics[width=0.5\linewidth]{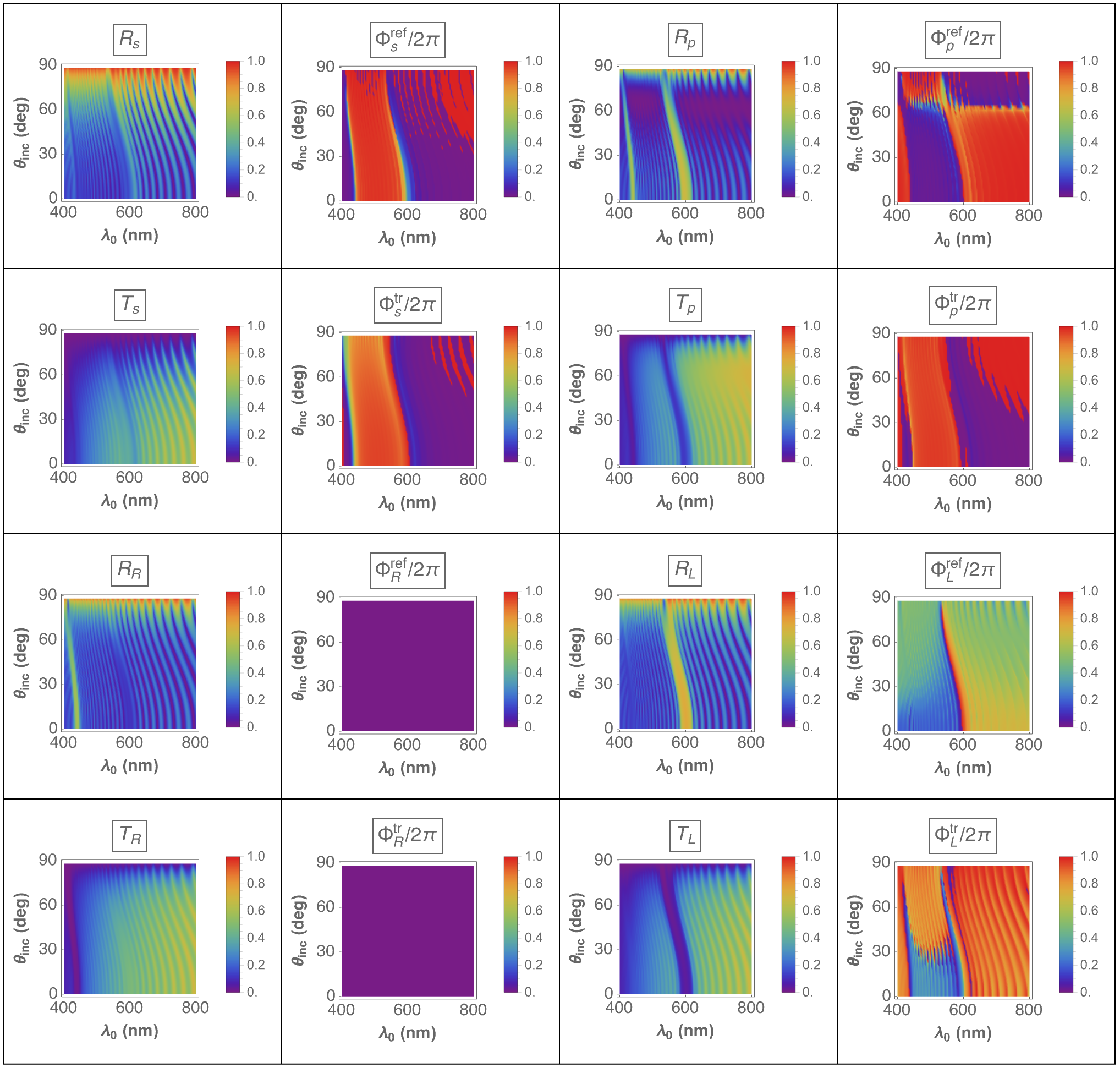}
	\end{center}
	\caption{Spectrums of $R_{\ell}$,   $T_{\ell}$, $\bPhiref_{\ell}$, and   $\bPhitra_{\ell}$, $\ell\in\left\{s,p,R,L\right\}$,
	for $\thetainc\in[0\deg,90\deg)$,
	when $\Nlyr=5$, $h= -1$, $N=15$,  $P=300$~nm, and $\psi=0\deg$.}
	\label{theta-Nlyr5-hneg}
\end{figure}

\begin{figure}[ht]
	\begin{center}
			\includegraphics[width=0.5\linewidth]{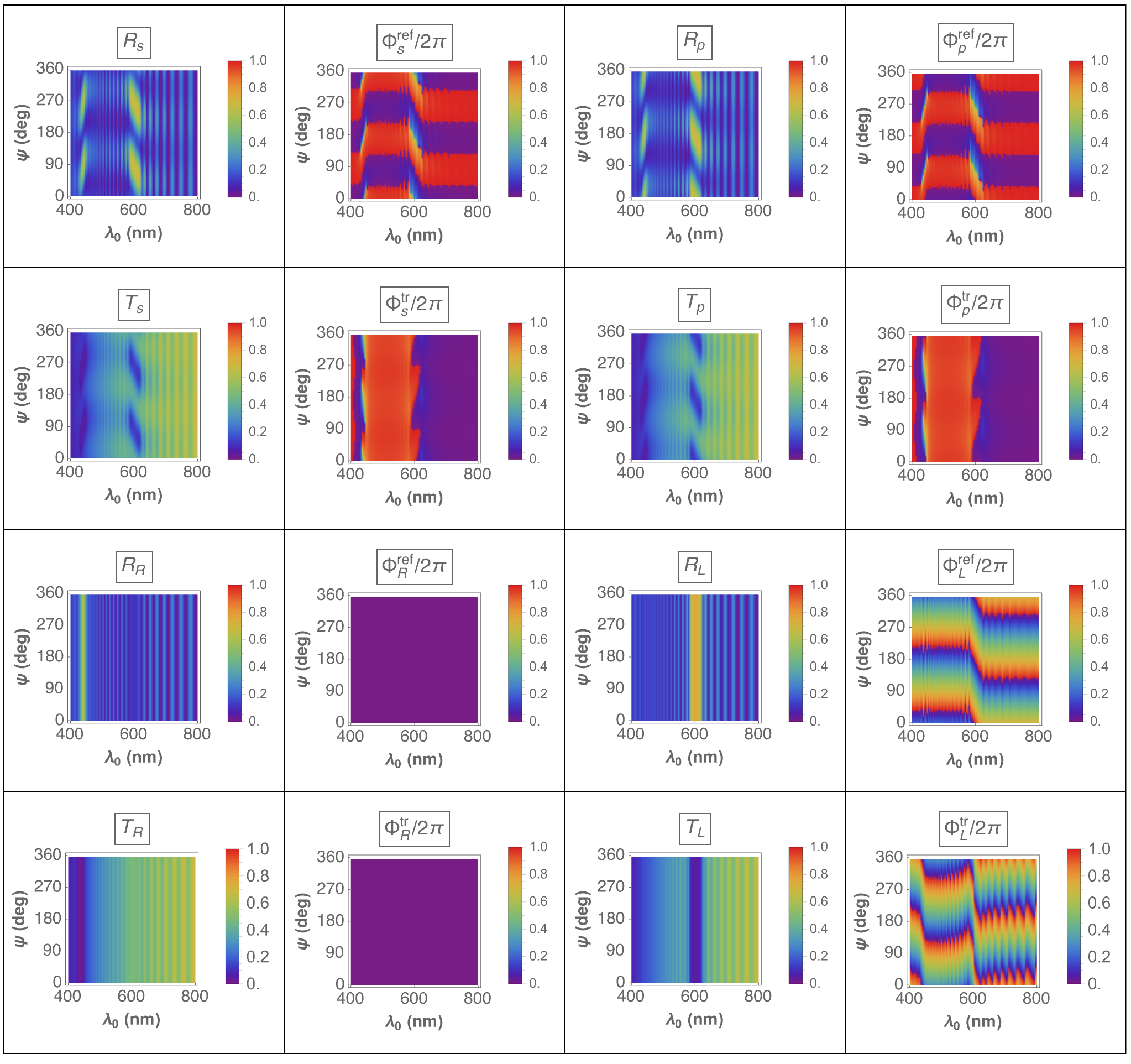}
	\end{center}
	\caption{Spectrums of $R_{\ell}$,   $T_{\ell}$, $\bPhiref_{\ell}$, and   $\bPhitra_{\ell}$, $\ell\in\left\{s,p,R,L\right\}$,
	for $\psi\in[0\deg,360\deg)$,
	when $\Nlyr=5$, $h= -1$, $N=15$,  $P=300$~nm, and $\thetainc=0\deg$.}
	\label{psi-Nlyr5-hneg}
\end{figure}

Figure~\ref{theta-Nlyr5-hpos} presents the chosen
spectrums for $\thetainc\in[0\deg,90\deg)$ and $\psi=0\deg$, and Fig.~\ref{psi-Nlyr5-hpos} for $\thetainc=0\deg$ and $\psi\in[0\deg,360\deg)$,
when $\Nlyr=5$ and $h=1$. Analogous spectrums for $\Nlyr=5$ and $h=-1$ are presented in Figs.~\ref{theta-Nlyr5-hneg} and \ref{psi-Nlyr5-hneg}. 
Regardless of whether $h=1$ or $h=-1$, two distinct circular Bragg regimes are evident in these plots. Whereas $\lambdao^{1;0,5/2}=602$~nm for
$\thetainc=0\deg$,
we have $\lambdao^{2;0,5/2}=439$~nm for normal incidence, so that $\lambdao^{2;0,5/2}/\lambdao^{1;0,5/2}=0.729$. This ratio is smaller than $0.667$ predicted 
by Eq.~(\ref{seq2}),  because the relative permittivity scalars in
Eq.~(\ref{resonance}) are frequency dependent.

Equations (\ref{circ-symm}) and (\ref{lin-symm}) hold, but  
similar symmetries are not evident for the non-zero geometric phases in Figs.~\ref{theta-Nlyr5-hpos}--\ref{psi-Nlyr5-hneg}.
Furthermore,  the spectral variations of $\bPhiref_{\ell}$ and   $\bPhitra_{\ell}$, $\ell\in\left\{s,p,L\right\}$,
are different in the two circular Bragg regimes in these figures.

\begin{figure}[!htb]
\begin{subfigure}{0.2\textwidth}
\centering
\includegraphics[width=1.8cm]{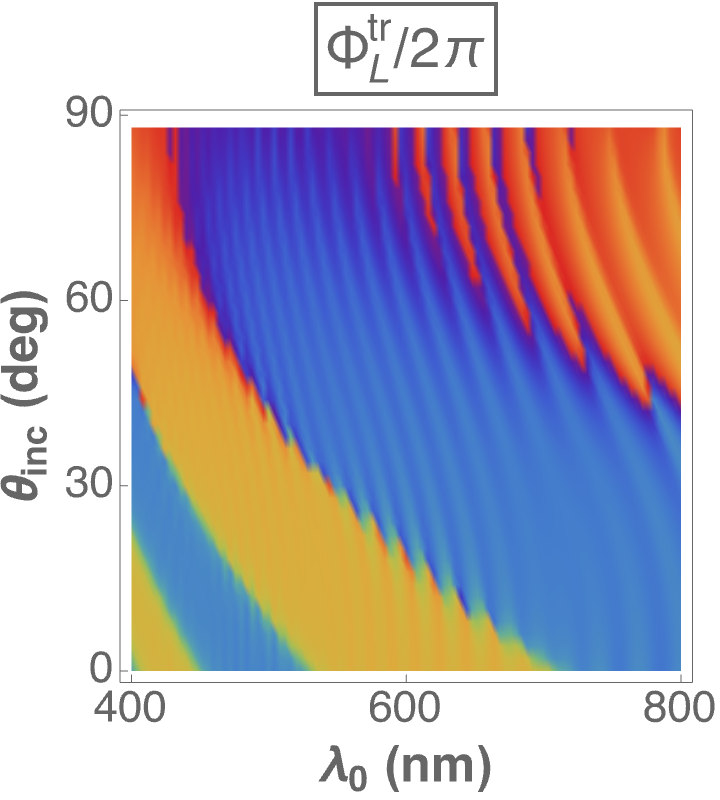} 
\hfill \vspace{0mm} 
 \caption{\label{Fig20-1} $\Nlyr=1$}
\end{subfigure}  \hspace{-5mm} 
\begin{subfigure}{0.2\textwidth}
\centering
\includegraphics[width=1.8cm]{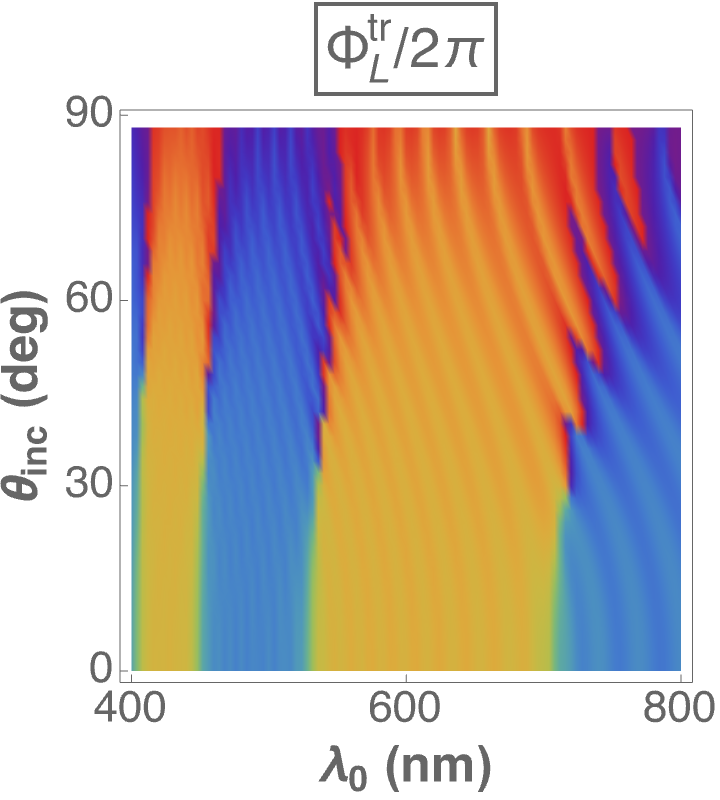} 
\hfill \vspace{0mm} 
 \caption{\label{Fig20-2} $\Nlyr=2$}
\end{subfigure} \hspace{-5mm}
\begin{subfigure}{0.2\textwidth}
\centering
\includegraphics[width=1.8cm]{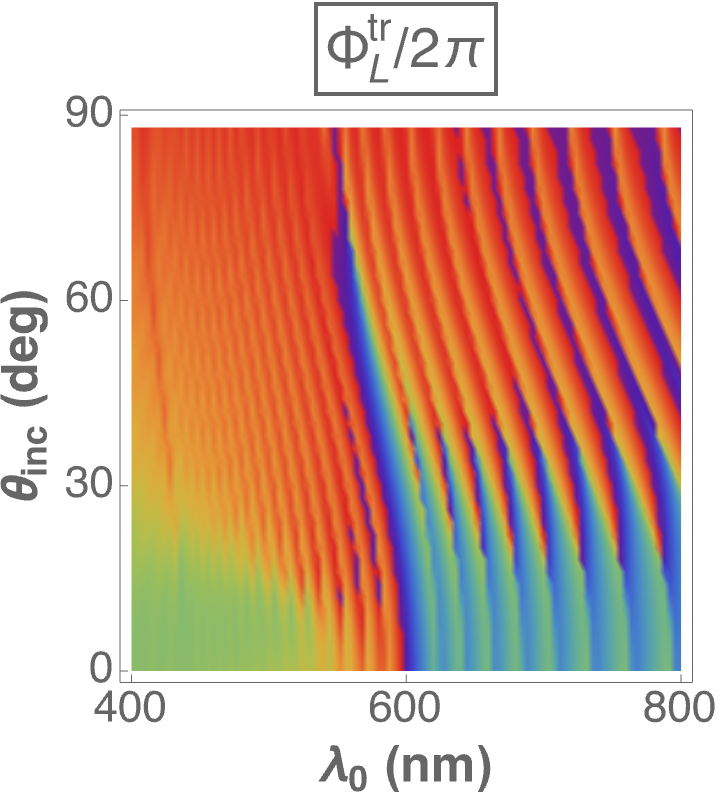} 
\hfill \vspace{0mm} 
 \caption{\label{Fig20-3} $\Nlyr=3$}
\end{subfigure} \hspace{-5mm}
\begin{subfigure}{0.2\textwidth}
\centering
\includegraphics[width=1.8cm]{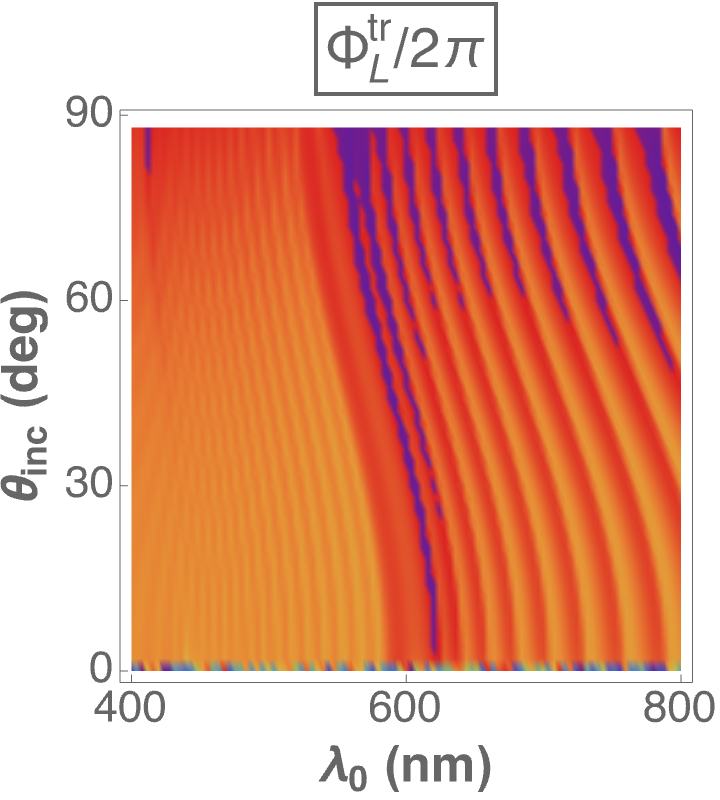} 
\hfill \vspace{0mm} 
 \caption{\label{Fig20-4} $\Nlyr=4$}
\end{subfigure} \hspace{-5mm}
\begin{subfigure}{0.2\textwidth}
\centering
\includegraphics[width=1.8cm]{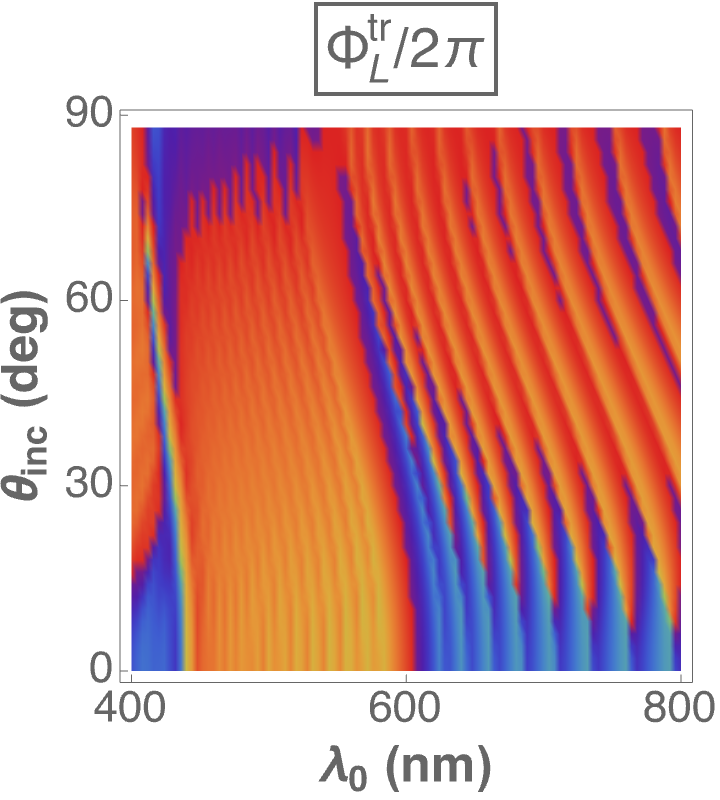} 
 \hfill \vspace{0mm} 
 \caption{\label{Fig20-5} $\Nlyr=5$}
\end{subfigure}\hspace{-3mm}
\begin{subfigure}{0.05\textwidth}
\centering
\includegraphics[width=0.45cm]{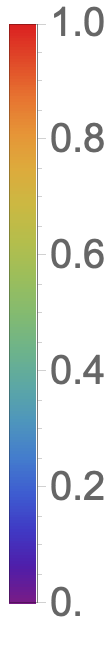} 
 \hfill \vspace{0mm} 
\end{subfigure}
\\
\begin{subfigure}{0.2\textwidth}
\centering
\includegraphics[width=1.8cm]{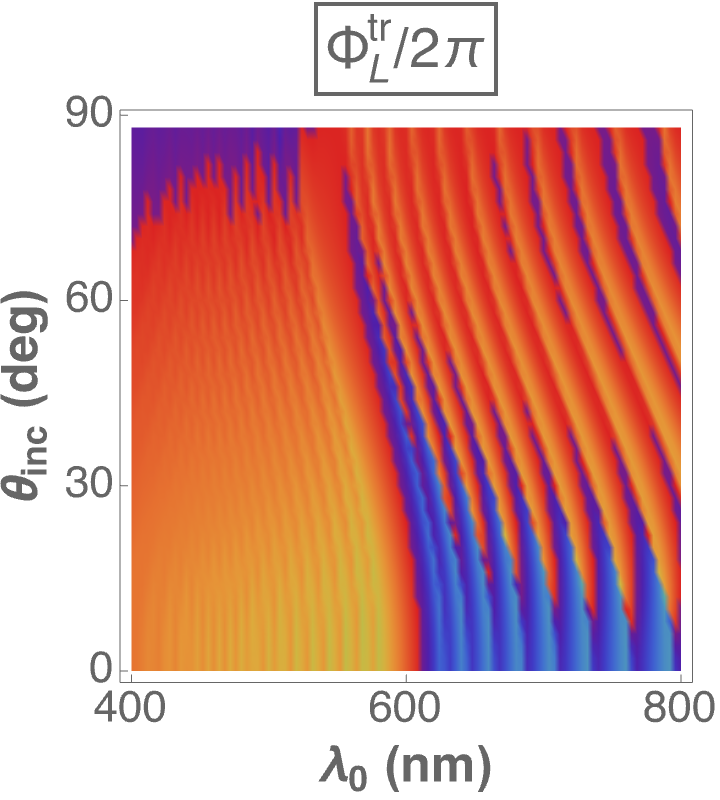} 
\hfill \vspace{0mm} 
 \caption{\label{Fig20-6} $\Nlyr=6$}
\end{subfigure}  \hspace{-5mm} 
\begin{subfigure}{0.2\textwidth}
\centering
\includegraphics[width=1.8cm]{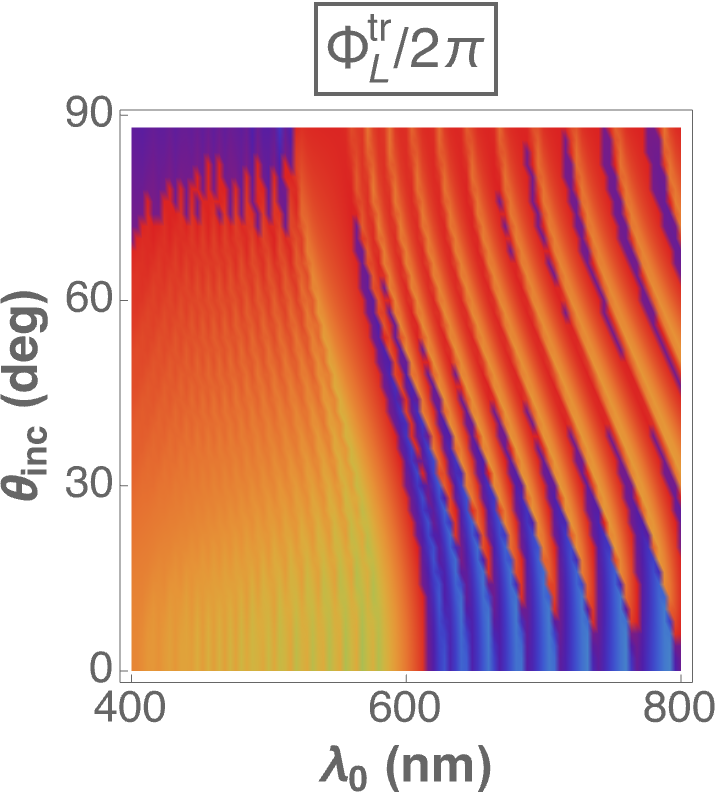} 
\hfill \vspace{0mm} 
 \caption{\label{Fig20-8} $\Nlyr=8$}
\end{subfigure} \hspace{-5mm}
\begin{subfigure}{0.2\textwidth}
\centering
\includegraphics[width=1.8cm]{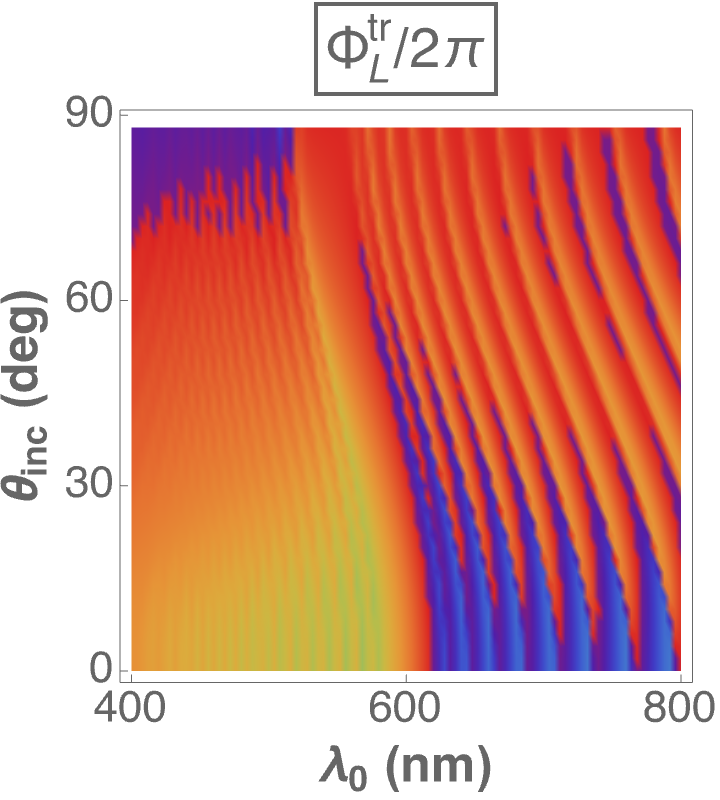} 
\hfill \vspace{0mm} 
 \caption{\label{Fig20-10} $\Nlyr=10$}
\end{subfigure} \hspace{-5mm}
\begin{subfigure}{0.2\textwidth}
\centering
\includegraphics[width=1.8cm]{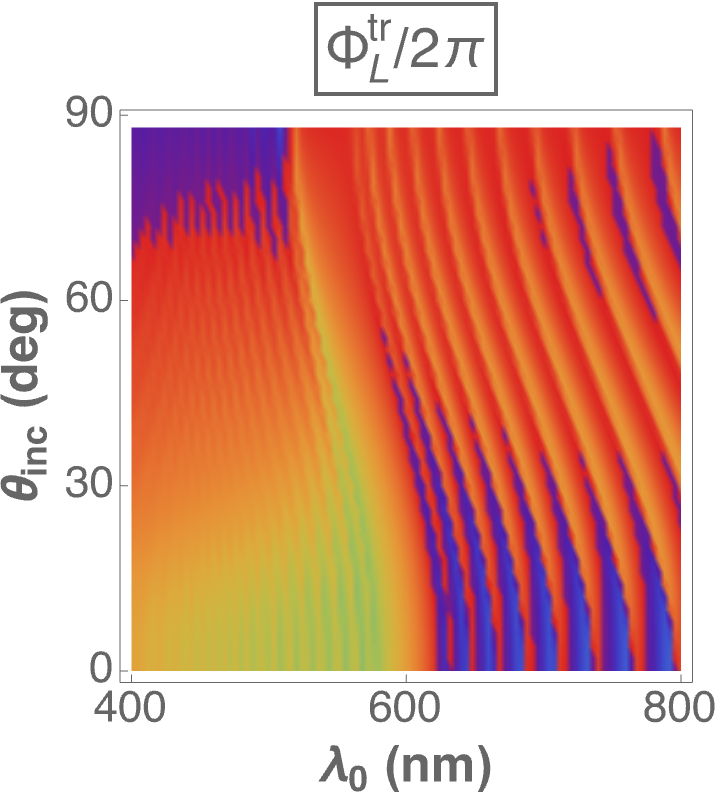} 
\hfill \vspace{0mm} 
 \caption{\label{Fig20-20} $\Nlyr=20$}
\end{subfigure} \hspace{-5mm}
\begin{subfigure}{0.2\textwidth}
\centering
\includegraphics[width=1.8cm]{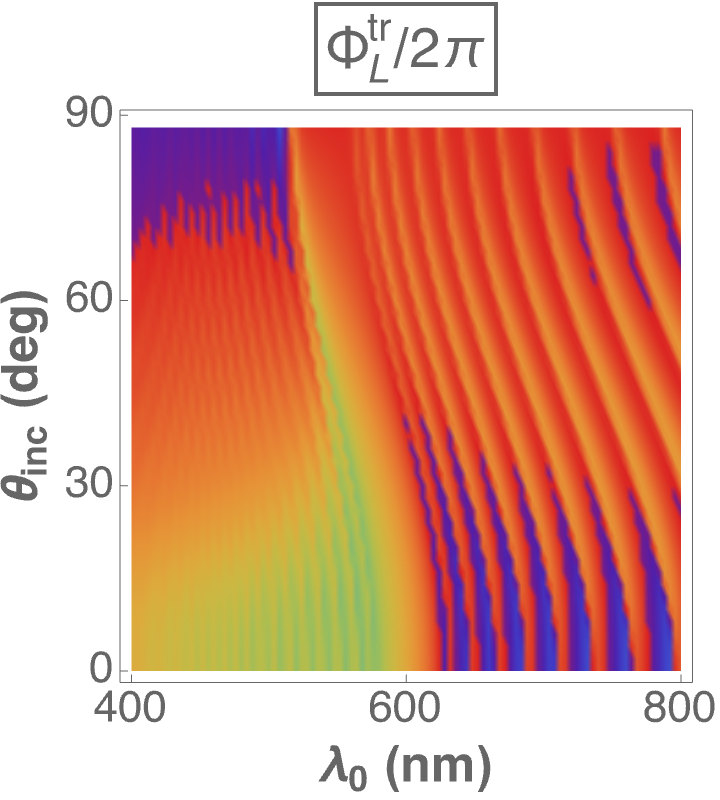} 
 \hfill \vspace{0mm} 
 \caption{\label{Fig20-50} $\Nlyr=50$}
\end{subfigure}\hspace{-3mm}
\begin{subfigure}{0.05\textwidth}
\centering
\includegraphics[width=0.45cm]{Lakhtakia_24_2hposScaleBar} 
 \hfill \vspace{0mm} 
\end{subfigure}
\\
\caption{Spectrums of $\bPhitra_{L}$ 
	for $\thetainc\in[0\deg,90\deg)$ and $\Nlyr\in\lec1,2,3,4,5,6,8,10,20,50\ric$,
	when $h= 1$, $N=15$,  $P=300$~nm, and $\psi=0\deg$.}
	\label{Comparison-hpos}
\end{figure}

\begin{figure}[!htb]
\begin{subfigure}{0.2\textwidth}
\centering
\includegraphics[width=1.8cm]{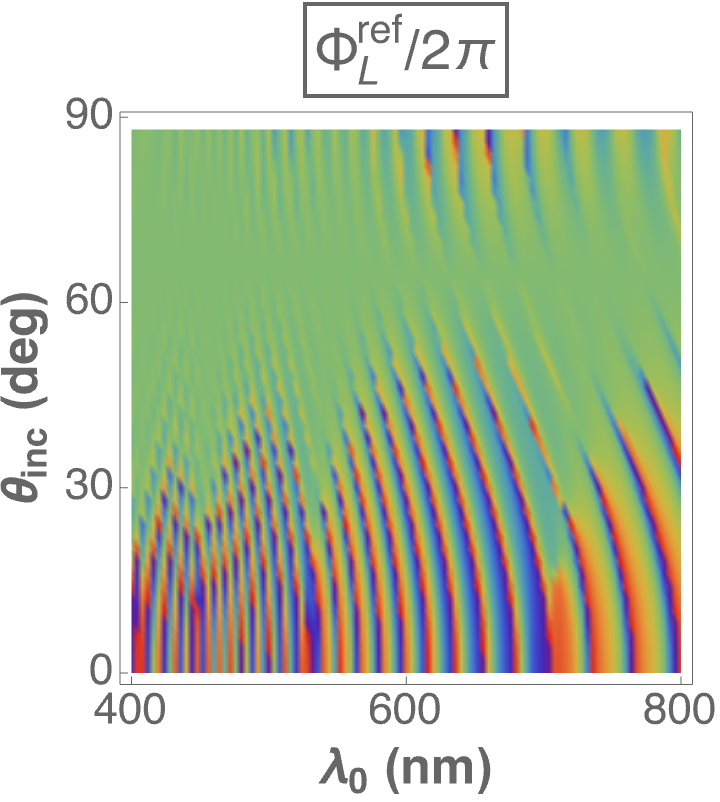} 
\hfill \vspace{0mm} 
 \caption{\label{Fig21-1} $\Nlyr=1$}
\end{subfigure}  \hspace{-5mm} 
\begin{subfigure}{0.2\textwidth}
\centering
\includegraphics[width=1.8cm]{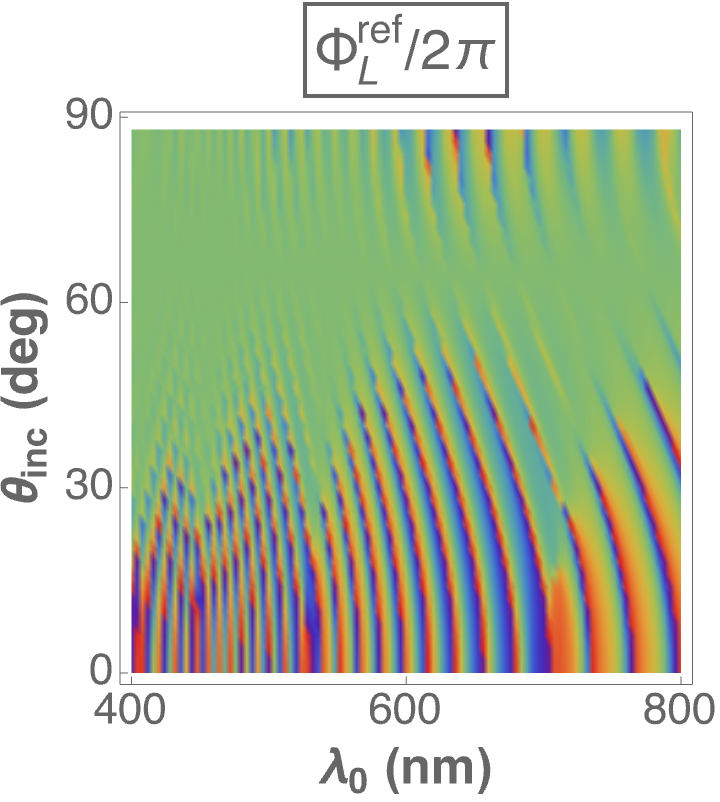} 
\hfill \vspace{0mm} 
 \caption{\label{Fig21-2} $\Nlyr=2$}
\end{subfigure} \hspace{-5mm}
\begin{subfigure}{0.2\textwidth}
\centering
\includegraphics[width=1.8cm]{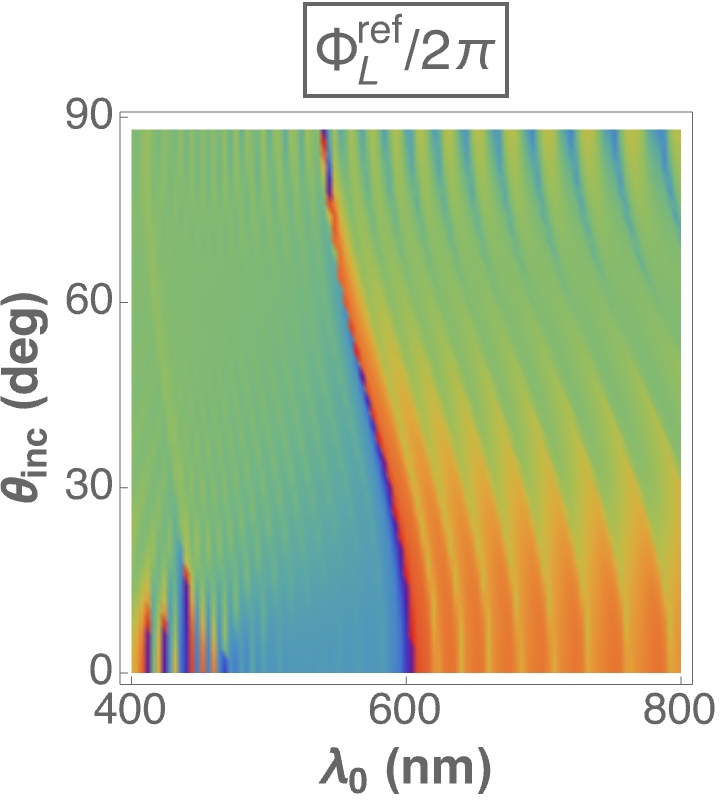} 
\hfill \vspace{0mm} 
 \caption{\label{Fig21-3} $\Nlyr=3$}
\end{subfigure} \hspace{-5mm}
\begin{subfigure}{0.2\textwidth}
\centering
\includegraphics[width=1.8cm]{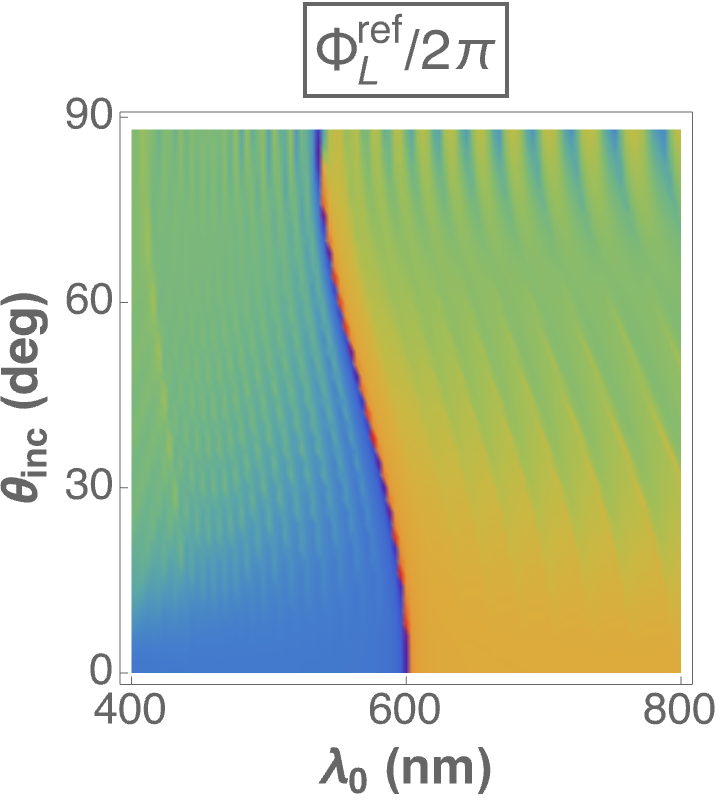} 
\hfill \vspace{0mm} 
 \caption{\label{Fig21-4} $\Nlyr=4$}
\end{subfigure} \hspace{-5mm}
\begin{subfigure}{0.2\textwidth}
\centering
\includegraphics[width=1.8cm]{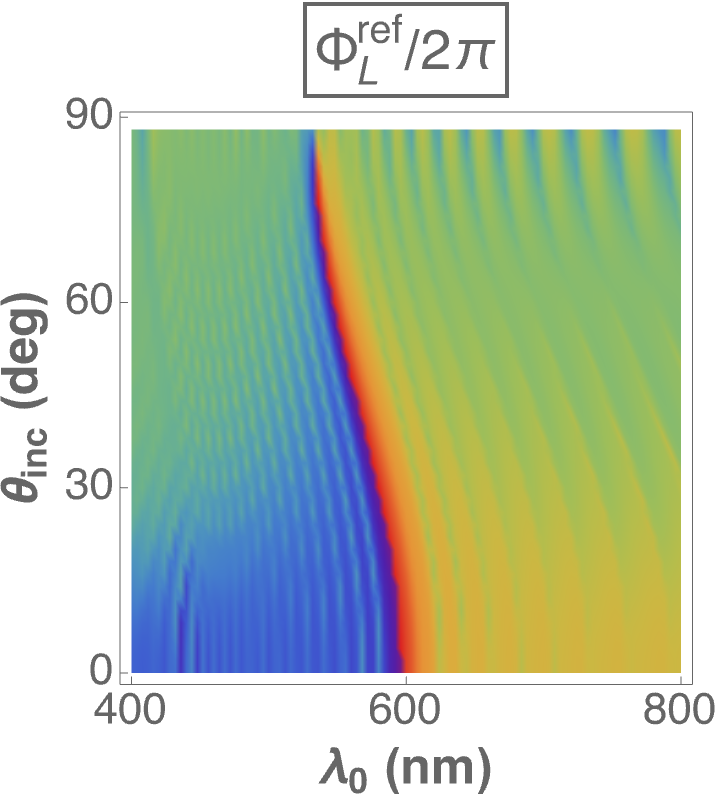} 
 \hfill \vspace{0mm} 
 \caption{\label{Fig21-5} $\Nlyr=5$}
\end{subfigure}\hspace{-3mm}
\begin{subfigure}{0.05\textwidth}
\centering
\includegraphics[width=0.45cm]{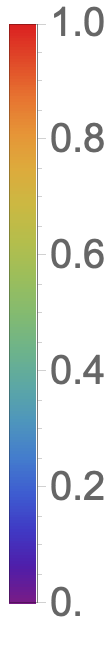} 
 \hfill \vspace{0mm} 
\end{subfigure}
\\
\begin{subfigure}{0.2\textwidth}
\centering
\includegraphics[width=1.8cm]{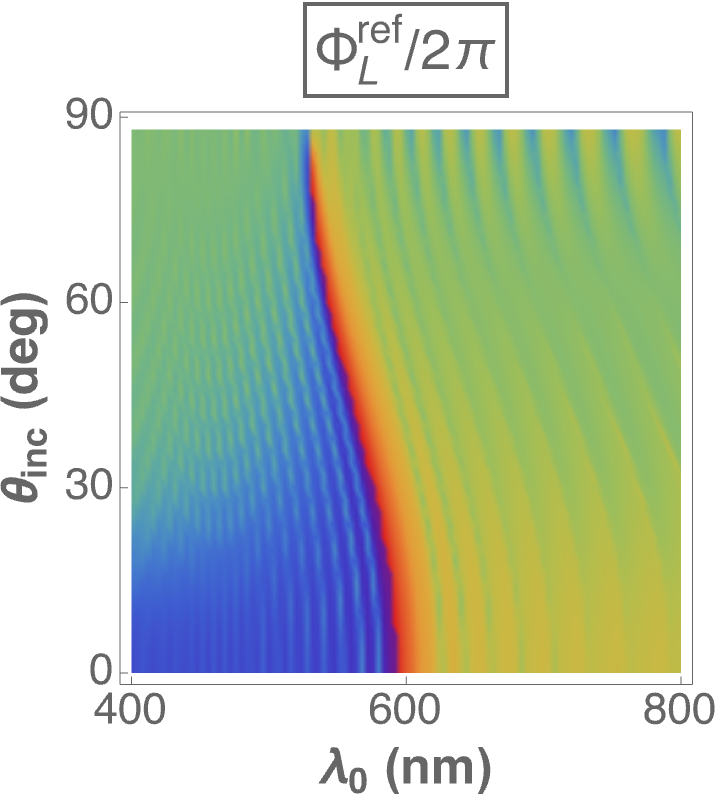} 
\hfill \vspace{0mm} 
 \caption{\label{Fig21-6} $\Nlyr=6$}
\end{subfigure}  \hspace{-5mm} 
\begin{subfigure}{0.2\textwidth}
\centering
\includegraphics[width=1.8cm]{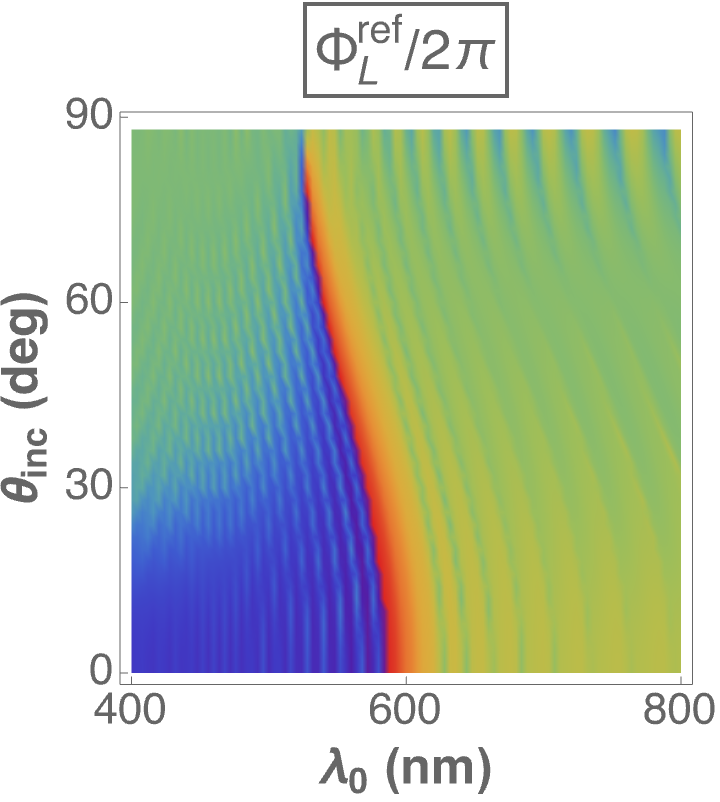} 
\hfill \vspace{0mm} 
 \caption{\label{Fig21-8} $\Nlyr=8$}
\end{subfigure} \hspace{-5mm}
\begin{subfigure}{0.2\textwidth}
\centering
\includegraphics[width=1.8cm]{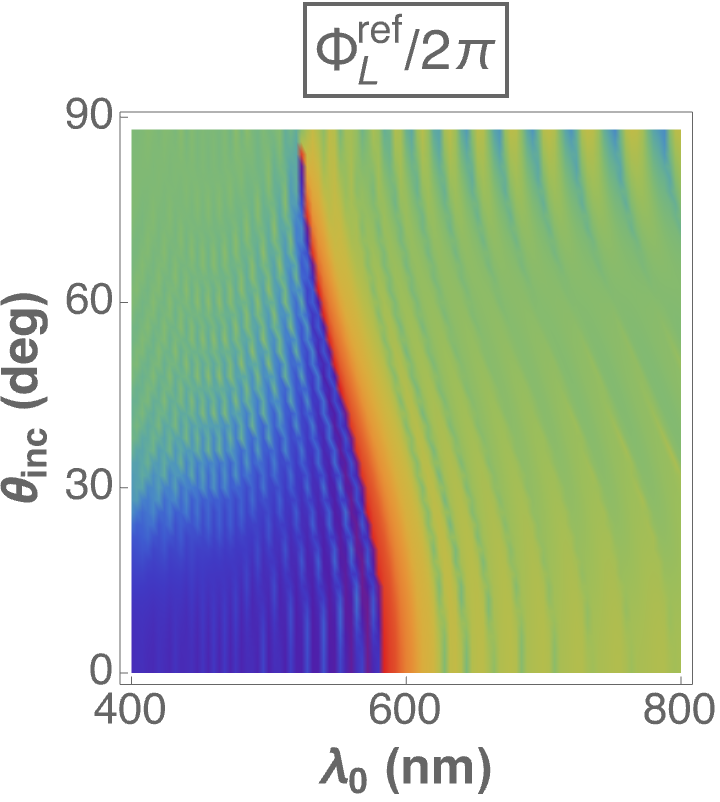} 
\hfill \vspace{0mm} 
 \caption{\label{Fig21-10} $\Nlyr=10$}
\end{subfigure} \hspace{-5mm}
\begin{subfigure}{0.2\textwidth}
\centering
\includegraphics[width=1.8cm]{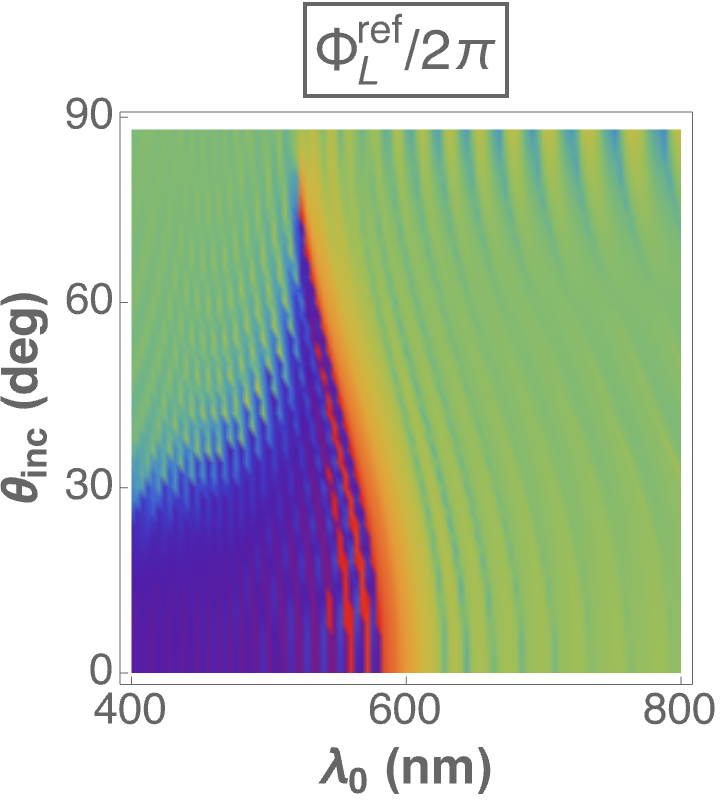} 
\hfill \vspace{0mm} 
 \caption{\label{Fig21-20} $\Nlyr=20$}
\end{subfigure} \hspace{-5mm}
\begin{subfigure}{0.2\textwidth}
\centering
\includegraphics[width=1.8cm]{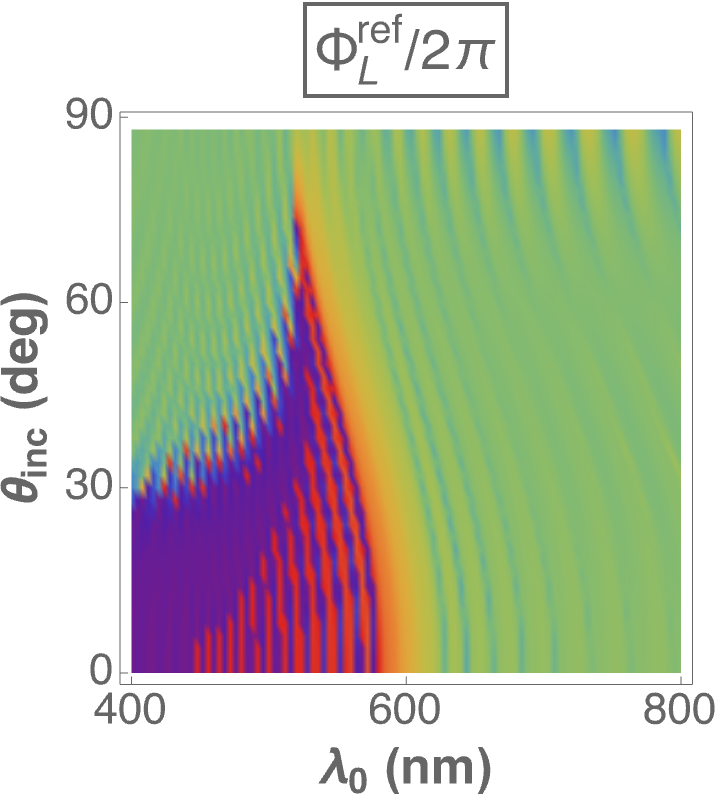} 
 \hfill \vspace{0mm} 
 \caption{\label{Fig21-50} $\Nlyr=50$}
\end{subfigure}\hspace{-3mm}
\begin{subfigure}{0.05\textwidth}
\centering
\includegraphics[width=0.45cm]{Lakhtakia_24_2hnegScaleBar} 
 \hfill \vspace{0mm} 
\end{subfigure}
\\
\caption{Spectrums of $\bPhiref_{L}$ 
	for $\thetainc\in[0\deg,90\deg)$ and $\Nlyr\in\lec1,2,3,4,5,6,8,10,20,50\ric$,
	when $h= -1$, $N=15$,  $P=300$~nm, and $\psi=0\deg$.}
	\label{Comparison-hneg}
\end{figure}

The Reusch pile becomes finely chiral and
$\lambdao^{2;0,\Nlyr/2}$ blueshifts ever farther from $\lambdao^{1;0,\Nlyr/2}$,
 as $\Nlyr$ increases significantly beyond $5$, and the   circular total remittance spectrums begin to converge
\cite{Abdulhalim2008}. A similar convergence is also exhibited by the linear total remittance spectrums. Spectrums of the non-zero
geometric phases also appear to converge, except in the vicinity of a circular Bragg regime. This becomes clear from examining the
plots of $\bPhitraL$ for $\psi=0\deg$, $h=1$ and $\Nlyr\in[1,50]$ in Fig.~(\ref{Comparison-hpos}), wherein the choice of $\bPhitraL$ over other
non-zero geometric phases takes into account the fact that $\TL$ is high inside the circular Bragg regime with $\lambdao^{1;0,\Nlyr/2}$ as
its center wavelength for $h=1$. The same conclusion emerges even more strongly  from examining the
plots of $\bPhirefL$ for $\psi=0\deg$, $h=-1$, and $\Nlyr\in[1,50]$ in Fig.~(\ref{Comparison-hneg}), the choice of $\bPhirefL$ over other
non-zero geometric phases taking into account that $\RL$ is high inside the circular Bragg regime with $\lambdao^{1;0,\Nlyr/2}$ as
its center wavelength for $h=-1$.

\section{Concluding remarks}\label{sec:cr}
Figures~(\ref{theta-Nlyr1})--(\ref{psi-Nlyr5-hneg}) provide an evolutionary perspective on the nanoscale architected morphology of Reusch piles.
Total remittances and geometric phases were presented as functions of the free-space wavelength and the direction
of plane-wave incidence for fixed thickness $NP=4500$~nm with $N=15$. As $\Nlyr$ increased from unity, first
the Reusch pile became structurally periodic (for $\Nlyr\geq2$) and then structurally handed (for $\Nlyr\geq3$). The layer thickness
$P/\Nlyr$ continued to shrink with increasing $\Nlyr$, which can be implemented quite straightforwardly with chiral 
sculptured thin films \cite{ambi,vanPopta}.  Total-remittance values in excess of $0.01$
began to converge  correct to four decimal digits, as $\Nlyr$
increased beyond $10$ (results not shown).

But that did not turn out to be completely true for the geometric phases. With $\bPhitraL$ selected for $h=1$ because $\TL$ has a substantial magnitude 
 inside the circular Bragg regime with $\lambdao^{1;0,\Nlyr/2}$ as
its center wavelength, and  $\bPhirefL$ selected for $h=-1$ because $\RL$ has a substantial magnitude inside the
same circular Bragg regime, to facilitate measurements eventually, Figs.~(\ref{Comparison-hpos}) and  (\ref{Comparison-hneg})
show the emergence of new features in the spectrums of non-zero geometric phases on the short-wavelength side
of the chosen circular Bragg regime. Since adjacent layers in a Reusch pile with high $\Nlyr$ are, at least theoretically identical
except for a slight twist about the $z$ axis, the foregoing observation implies that the geometric phases are sensitive to nanoscale
morphological changes as a thin film grows. Therefore, geometric phases could contain signatures of nanoscale morphological details that would depend
on the fabrication technique. Careful experiments are necessary to establish this possibility.

 \section*{Acknowledgment}
The author thanks the Charles Godfrey Binder Endowment at Penn State for supporting  his research from 2006 to 2024.

\end{document}